\documentclass[aps,preprint,groupedaddress]{revtex4-1} 
\usepackage{graphicx}
\usepackage{xcolor}
\usepackage[toc,nonumberlist]{glossaries}
\usepackage[capitalise]{cleveref}
\usepackage{booktabs}
\usepackage{amsfonts}
\usepackage{enumitem}
\usepackage{nicefrac}
\usepackage{fontawesome}
\usepackage{tabularx}
\usepackage{multirow}
\usepackage{makecell}
\usepackage{bm}

\newcolumntype{C}{>{\centering\arraybackslash}X}
\newcolumntype{f}[1]{>{\centering\arraybackslash}p{#1}}

\newcommand\given[1][]{\ensuremath{#1\mid}}

\newcommand{\vecb}[1]{\ensuremath{\boldsymbol{#1}}}

\usepackage[colorinlistoftodos]{todonotes}

\newacronym{mle}{MLE}{Maximum Likelihood Estimation}
\newacronym{jhep}{JHEP}{The Journal of High Energy Physics} \newacronym{jpr_ins}{PR-HEP}{High Energy Physics in Physical Review Journals} \newacronym{jpl}{Phys.\,Lett.}{Physics Letters A, B} \newacronym{jnp}{Nuc.\,Phys.}{Nuclear Physics} 
\newacronym{aps}{APS}{American Physical Society} 

\newacronym{jpr}{PR}{Physical Review} \newacronym{jpra}{PRA}{Physical Review A} \newacronym{jprb}{PRB}{Physical Review B} \newacronym{jprc}{PRC}{Physical Review C} \newacronym{jprd}{PRD}{Physical Review D} \newacronym{jpre}{PRE}{Physical Review E} \newacronym{jpri}{PRI}{Physical Review I} \newacronym{jprl}{PRL}{Physical Review L} \newacronym{jprstab}{PRSTAB}{Physical Review STAB}  \newacronym{jprstrep}{PRSTREP}{Physical Review STREP}  \newacronym{jrmp}{RMP}{Reviews of Modern Physics}

\newacronym{p320}{PAT\,320}{Patent Class 320} \newacronym{p424}{PAT\,424}{Patent Class 424} \newacronym{p703}{PAT\,703}{Patent Class 703}

\newacronym{uspc}{USPC}{United States Patent Classification}
\newacronym{uspto}{USPTO}{United States Patent and Trademark Office}

\newglossaryentry{ckn}
{
name=collaborative knowledge network,
description={two-mode network with one type of node representing knowledge artifacts, the other one representing persons, and one ``authorship'' type edges connecting persons and artifacts and directed ``citation'' type edges connecting artifacts}
}

\AtBeginDocument{
	\heavyrulewidth=.08em
	\lightrulewidth=.05em
	\cmidrulewidth=.03em
	\belowrulesep=.65ex
	\belowbottomsep=0pt
	\aboverulesep=.4ex
	\abovetopsep=0pt
	\cmidrulesep=\doublerulesep
	\cmidrulekern=.5em
	\defaultaddspace=.5em
}

\usepackage{tikz}

\definecolor{cpaper}{HTML}{004F80}
\definecolor{cauthor}{HTML}{f0c304}
\definecolor{ctext}{HTML}{000000}

\usetikzlibrary{calc,3d}
\usetikzlibrary{decorations.pathreplacing}

\tikzstyle{article}=[draw,circle,white,fill=cpaper, inner sep=0pt, minimum width=2.5mm]
\tikzstyle{author}=[draw,circle,black,fill=cauthor, inner sep=0pt, minimum width=2.5mm]
\tikzstyle{cit}=[draw,cpaper,line width=0.4mm]
\tikzstyle{collab}=[draw,cauthor,line width=0.4mm]

\tikzset{
    position/.style args={#1:#2 from #3}{
        at=(#3.#1), anchor=#1+180, shift=(#1:#2)
    }
}

\usepackage{mdwlist}

\renewenvironment{description}{
\begin{basedescript}{
\desclabelstyle{\nextlinelabel}
\desclabelwidth{2em}}}
{
\end{basedescript}
}

\begin{document}

 \title{A multi-layer network approach to modelling authorship influence on citation dynamics in physics journals}

\author{Vahan Nanumyan}
\email[]{vnanumyan@ethz.ch}
\author{Christoph Gote}
\email[]{cgote@ethz.ch}
\author{Frank Schweitzer}
\email[]{fschweitzer@ethz.ch}
\affiliation{Chair of Systems Design, ETH Zurich, Weinbergstrasse 58, 8092 Zurich, Switzerland}

\date{\today}

\begin{abstract}
  We provide a general framework to model the growth of networks consisting of different coupled layers.
  Our aim is to estimate the impact of one such layer on the dynamics of the others. 
  As an application, we study a scientometric network, where one layer consists of publications as nodes and citations as links, whereas the second layer represents the authors.
  This allows to address the question how characteristics of authors, such as their number of publications or number of previous co-authors, impacts the citation dynamics of a new publication.
  To test different hypotheses about this impact, our model combines citation constituents and social constituents in different ways.
  We then evaluate their performance in reproducing the citation dynamics in nine different physics journals.
  For this, we develop a general  method for statistical parameter estimation and model selection that is applicable to growing multi-layer networks.
  It takes both the parameter errors and the model complexity into account and is computationally efficient and scalable to large networks.
    \end{abstract}


\maketitle

\graphicspath{{./figures/}}

\section{Introduction}\label{sec:intro}

Citation networks and their growth have attracted the interest of the complex networks research community for long~\cite{wang2009effect,Parolo2015,Golosovsky2017}.
A citation network is a directed network, in which nodes represent scientific publications and edges between nodes citations between publications.
Because new publications can only cite existing ones, the \emph{temporal ordering} in which nodes and links are added to the network plays a crucial role, and a time-aggregated network representation omits this information. 

One focus of current research was to test whether established growth mechanisms, such as preferential attachment, are suitable to describe the evolution of citation networks~\cite{wang2008measuring,Golosovsky2013,Golosovsky2018}.
Preferential attachment assumes the growth rate, \(\Delta k\), is proportional to the in-degree of a node, $k$, i.e. the number of citations a paper has attracted so far. 
This follows early investigations from the 1960's showing that publications with already a large number of citations tend to attract even more citations~\cite{Price1965,Price1976}.

Preferential attachment prescribes a certain average dynamics to every node, i.e. it does not account for any kind of \emph{heterogeneity} other than the degree.
In practice, however, some nodes, once added to the network, accumulate many edges very fast, while others do not.
This observation has led to the notion of node \emph{fitness}: the fitter a node, the more edges it attracts~\cite{Bianconi2001,Wang2013,PhysRevE.99.060301}.
Conceptually, this considers a heterogeneity among nodes that goes beyond network properties. 
In citation networks, the fitness is usually attributed to the content of the paper.

Preferential attachment further implies that that \emph{older} nodes attract more edges, which is in disagreement with empirical observations of citation dynamics. 
Hence, the \emph{ageing} of nodes was introduced as another node property.
With respect to publications, ageing reflects obsolescence (novelty decay, relevance decay)~\cite{Wu2007,Medo2011} or attention decay~\cite{Parolo2015,Schweitzer2019}.

When verifying the preferential attachment hypothesis, most works either looked at the scale-free \emph{degree distribution} as an aggregated property of the network, or analysed the relation between in-degree, $k$, and growth rate, \(\Delta k\).
These approaches suffer from certain drawbacks.
For instance, the mentioned distributions and relations can result from many dynamic mechanisms, e.g.  \(\Delta k \sim k\) can be found from a  preferential attachment or a fitness model~\cite{Golosovsky2018}.
In addition, results on statistical modelling are not conclusive without the statistical comparison with other candidate models and without calculating the parameter errors.
For example, without errors or confidence intervals for the exponent of the degree distribution in the non-linear preferential attachment model, the discussion about evidence for or against a non-linearity in the growth mechanism is not complete~\cite{Golosovsky2013}.

In addition to the methodological issues with applying the preferential attachment model to citation growth, there are also conceptual problems.
The most important one regards the role of social influences on the citation dynamics.
Specifically, to what extent do properties of the \emph{authors}, e.g. their reputation as expressed by their number of previous publications or their total number of citations, play a role in citing a publication?
After all, producing academic publications is a \emph{social endeavour} that involves collaboration between authors and information spreading through social interactions.

Thus, it is reasonable to expect that these social aspects have an impact on  the citation dynamics.
Already \citet{Merton1968} in his seminal work discussed various mechanisms of how characteristics of authors (prominence, academic awards) may influence the recognition of their publications in terms of citations.
In particular, he raised the question of whether publications by better known researchers get more and faster recognition in the community.

To formally address such issues in a modeling approach requires us to consider not only the relations between \emph{publications}, in a citation network, but additionally also the relations between \emph{authors}, in a co-authorship network.
Such collaboration networks between authors~\cite{Tomasello2017} and their co-evolution with the citation networks~\cite{Boerner2004}  have been studied recently. 
For instance, it was shown that the centrality of authors in the co-authorship network prior to a publication affects the number of citations that the publication will receive~\cite{Sarigol2014}, which allows to predict their future success.

In this article, we provide an in-depth analysis to quantify the impact of authors' characteristics, such as their previous number of publications or co-authors, on the citation growth of their publications.
Specifically, we will test different hypotheses about this impact and combine them with different growth mechanisms for the citation dynamics.
This way, our results shed a new light on the discussion about the ``fitness'' of a publication.
While it is usually attributed to the \emph{content} of the paper, we address the question whether this ``fitness'' can be better attributed to the properties of the \emph{authors}.
This would allow to transform the rather abstract notion of ``fitness'' in relation to ``content'' into a measurable and interpretable quantity.

To achieve our goal, we start from the more general perspective of coupled multi-layer networks.
We provide a framework to model the growth of such coupled networks, but even more, we demonstrate how a statistical evaluation of such growth models should be carried out. 
This generalises the methods presented in \cite{Leskovec2008,Medo2014} for single layer networks.
We refine these methods in two ways.
First, we demonstrate how to compute the standard errors of the parameter estimates.
This allows us to judge the significance of the model parameters and the corresponding model formulations.
Second, we address the issue of prohibitive computational cost of the microscopic model evaluation based on the sequential addition of individual edges. 
We solve this for arbitrarily large networks by evaluating the models based on finite samples of growth events over the course of the network growth.
By sampling events uniformly over the whole period of network growth, from the first nodes to the final observed state of the network, we ensure representing the whole growth process.

The article is organised as follows.
In \cref{sec:multi-layer-growth} we formalise (i) the multi-layer network representation of citations between publications and authorship relations between publications and authors, (ii) the modelling approach to coupled growth of multi-layer networks in general.
In \cref{sec:temp-mle}, we introduce the maximum likelihood estimation of model parameters and model selection based on the temporal sequence of edges added to the network, the scalable estimation procedure based on event sampling, and a way to evaluate the temporal stability of model parameters.
In \cref{sec:valid-synth-data} we validate the methods on synthetic networks generated with known mechanisms.
We then proceed with the formulation of specific models for citation growth of publications coupled with the properties of the authors of these publications in \cref{sec:model-spec-citat}.
Finally in \cref{sec:scientometric-results}, we analyse the growth of nine empirical citation networks that have been constructed from physics journals based on data from the American Physical Society and INSPIRE.

\section{A multi-layer growth model}
\label{sec:multi-layer-growth}

\subsection{A multi-layer approach to citation dynamics}
\label{sec:multi-layer-approach}

The real-world application of our general framework comes from the domain of scientometrics.
Specifically, we want to identify the mechanisms that drive citations between different publications.
We therefore use data from nine physics journals, most of them from the APS, which are described in detail in \cref{sec:appendix_data}.
For each journal and each publication therein, we extract data about (i) the publication time, (ii) the authors and (iii) the  publications cited in the respective publication. 

\begin{figure}[htbp]
\begin{center}
  \begin{tikzpicture}[x  = {(0.5cm,0.5cm)},
						y  = {(1cm,0cm)},
						z  = {(0cm,1cm)},
						scale = 1,
						color = {ctext},
						every node/.append style={transform shape},
						label distance=-1mm,
						font=\sffamily]

	\tikzset{layerstyle/.style={fill opacity=.2, draw=black, draw opacity=.0, very thin, line join=round}}
	
	\begin{scope}[canvas is yx plane at z=0]
		\path[layerstyle,fill=cauthor] (0.5,1) rectangle (6,4.5);
		\node[anchor=south west] at (0.5,1) {authors};
		
		\foreach \place/\name in {{(1,3)/Alice}, {(2,2)/Bob}, {(2,4)/Carol}, {(4.2,2.7)/Dave}}
			\node[author,label=right:\name] (\name) at \place {};
	\end{scope}

	\foreach \place/\name in {{(3,2,2)/art1}, {(2,4,2)/art2}, {(3,4.5,2)/art4}, {(2,2,0)/au1}, {(4,2,0)/au2}, {(3,1,0)/au3}, {(2.7,4.2,0)/au4}, {(2,6,0)/au5}, {(4,6,0)/au6},{(3+0.71,4.5+0.71,2)/art3}}
	 \node[inner sep=0pt, minimum width=2mm] (\name) at \place {};
	\foreach \source/\dest in {art1/au1, art1/au2, art1/au3, art2/au2, art2/au4, art4/au4}
	 \path (\source) edge[cauthor,line width=0.4mm] (\dest);

	\begin{scope}[canvas is yx plane at z=2, every path/.append style={cit}]
		\path[layerstyle,fill=cpaper] (0.5,1) rectangle (6,5);
		\node[anchor=south west,ctext] at (0.5,1) {publications};
		
		\foreach \place/\name in {{(4,2)/a}, {(4.5,3)/b}, {(2.7,3.5)/d}, {(2,3)/e}}
			\node[article] (\name) at \place {};
		\foreach \source/\dest in {a/b, a/d, e/a, e/d}
			\path (\source) edge[-latex] (\dest);
		 %
		 \foreach \pos/\i in {above left of/1, left of/2, below left of/3}
			 \node[article, \pos = e] (e\i) {};
		 \foreach \speer/\peer in {e/e1,e/e2,e/e3}
			 \path (\speer) edge[latex-] (\peer);
		 \foreach \pos/\i in {right of/2, below right of/3}
			 \node[article, \pos =b ] (b\i) {};
		 \foreach \speer/\peer in {b/b2,b/b3}
			 \path (\speer) edge[latex-] (\peer);
		 \node[article, above of=d] (d1){};
		 \path (d) edge[latex-] (d1);
		 \foreach \pos/\i in {below left of/1, below right of/2}
			 \node[article, \pos =a ] (a\i) {};
		 \foreach \speer/\peer in {a/a1,a/a2}
		 \path (\speer) edge[latex-] (\peer);
	\end{scope}

	\end{tikzpicture}
\end{center}
\caption{Multilayer representation of authors and publications.}
	\label{fig:cit-aut-multi}
\end{figure}
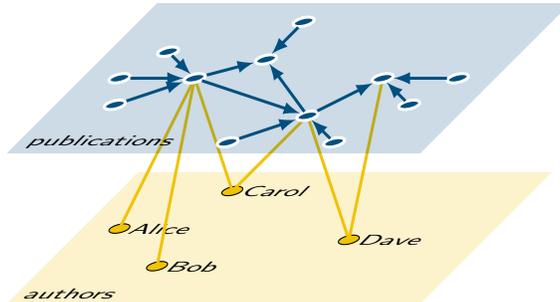

To represent this relational data, we construct a two-layer network (see Figure~\ref{fig:cit-aut-multi}) in which the lower layer consists of the authors and the upper layer of their publications. 
Edges in the upper layer indicate citations between publications.
The relation \emph{between} the two layers is defined by the co-authorship for each publication. 
Previous works have mainly focused on the dynamics of the publication layer, which was studied in isolation, i.e., without accounting for influences from the author layer~\cite{Price1965,Barabasi1999,wang2008measuring,wang2009effect,Wang2013,Parolo2015,Lariviere2008}.
This shall be improved in our paper. 
The research question addressed with respect to this two-layer representation can be stated as follows:
To what extent impacts the author layer (i.e. the status, collaboration history etc. of authors) the publication layer (i.e. citation of papers)?

\begin{figure}[htpb]
	\centering
	\includegraphics[width=.45\textwidth]{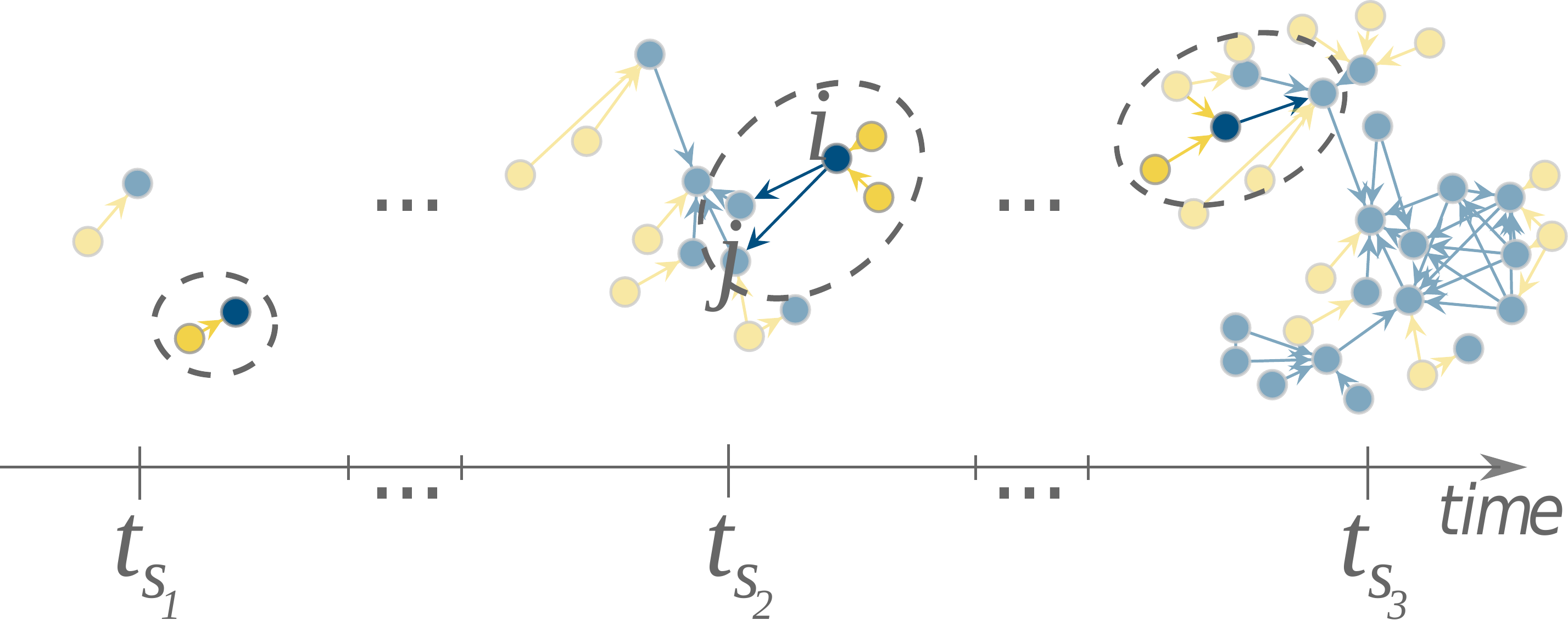}
	\caption{
          Coupled dynamics of the two-layer network. Blue nodes/edges refer to the publication layer, yellow nodes/edges to the author layer (see also Figure~\ref{fig:cit-aut-multi}). Each growth event is associated with a new publication, which draws citation edges to older publications and co-authorship edges between new and/or existing authors.}
	\label{fig:net-evol}
\end{figure}

While this research question is interesting in itself, it does not address the notion of \emph{time} in an explicit manner.
We have to take into account that publications enter the system at different times, which restricts the range of possible citations to papers already published.
With every new publication, also the author layer changes. 
Figure \ref{fig:net-evol} illustrates with three time-ordered snapshots such a coupled growth process.
To capture this dynamics properly, in this paper we formulate a \emph{growth} model for the publication layer that takes changes in the  author layer into account. 
The related research question can be stated as follows: What is the impact of time on these relations, specifically the decay of attention towards older publications?

\subsection{A multi-layer approach to  coupled growth processes}
\label{sec:model-general}

The application of multi-layer networks to scientometrics  is just one instance of a bigger class of problems, in which two processes, \emph{within} and \emph{across} different layers, concurrently determine the overall dynamics.
Therefore, in this paper we want to derive a general approach to these problems, before applying it to the specific application of the citation dynamics.

\begin{figure}[t]\centering
  \includegraphics[width=.48\textwidth]{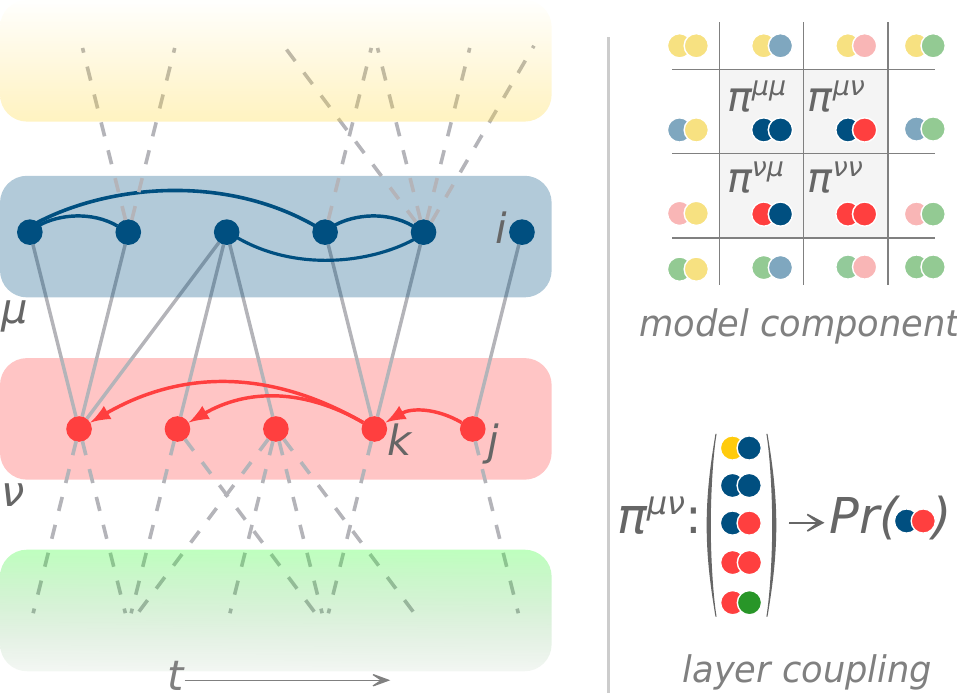}
  \caption{(Left) Multi-layer network representation. For illustrative purposes, only edges between neighbouring layers are drawn.
  (Right top) The growth of the network can be described by means of separate model components \(\pi^{\mu\nu}\), each corresponding to edge creation of one type, \(E^{\mu\nu}\). 
  (Right bottom) The coupling between the dynamics of different layers is captured in the model components.
  }\label{fig:multi-layer}
\end{figure}

We consider a growing multi-layer network \(G(t) = G\left[V(t),E(t)\right]\) with \(L\) sets of nodes \(V(t) = \{V^\mu(t)\}_{\mu \in [1,L] }\) that define the network layers \(\mu \in [1,L]\) and with edges \(E(t) = \{E^{\mu\nu}(t)\}_{(\mu,\nu) \in [1,L]\times[1,L]}\).
Each set of edges \(E^{\mu\nu}(t)\) comprises all and only edges that connect nodes in layer \(V^\mu(t)\) to nodes in \(V^\nu(t)\).
The left panel in \cref{fig:multi-layer} illustrates such a multi-layer network where the edges within and between two layers \(\mu\) and \(\nu\) shown.
In principle, the edges within and across different layers can be directed or undirected independently from each other.
Edges can also be weighted, but for simplicity we will not consider the weights in the further discussion.
This notation above is quite general and can encode the special cases of multiplex networks \citep{Nicosia2013} in which \(V^\mu(t) = V^\nu(t)\) for all \(\mu,\nu \in [1,L]\), or multi-partite networks \citep{garas16:_inter} in which \(E^{\mu\mu} = \emptyset\) for any \(\mu\) and \(\nu\).

In the following we investigate \emph{growing} networks, i.e. nodes and edges are added to the network over time. 
We do not consider isolated nodes and the removal of nodes or edges. 
Formally, our models define the probability of adding an \emph{edge} \(e_{ij} \in E^{\mu\nu}(t+1)\) to the network at time \(t+1\).
We assume discrete time, parametrised by the network growth.
This means that a time step \(t \to t+1\) occurs only when at least one edge (but potentially, multiple simultaneous edges) is added to the network.
The probability of this event is, in its most general form, expressed as: 
\begin{equation}
  \label{eq:general-model}
  \Pr \left[ e_{ij} \in E^{\mu\nu}(t+1) \land e_{ij} \notin E^{\mu\nu}(t) \right] = \pi^{\mu\nu}_{ij}\left[G(t);\vecb{\theta}\right].
\end{equation}
\(\pi^{\mu\nu}_{ij}\left[G(t);\vecb{\theta}\right]\), which is the central element of our model specifications below, defines this probability as a function of the network $G(t)$, taken at the time before the edge is added, and a vector of parameters, \(\vecb{\theta}\), which later allow to distinguish between different models.

Note that, in the right panel of \cref{fig:multi-layer}, the \(\pi^{\mu\nu}_{ij}\left[G(t);\vecb{\theta}\right]\) denote \emph{model components}. 
They may differ for different sets of edges and are therefore indexed by the network layers \(\mu\) and \(\nu\).
In principle, each set of edges \(E^{\mu\nu}\) can be modelled by its own model component.
Because each model component is in general a function of the whole network at time \(t\), the addition of edges in one network layer is affected by the state of the network not only in that layer but also in other layers.
In other words, this general formulation allows for coupling in the growth dynamics of different layers.
In Sect.~\ref{sec:valid-synth-data}, we will specify the \(\pi^{\mu\nu}_{ij}\left[G(t);\vecb{\theta}\right]\) to capture the growth of a synthetic network.
More importantly, in Sect.~\ref{sec:model-component} we will present a general expression for the model component \(\pi^{pp}_{ij}\left[G(t);\vecb{\theta}\right]\) that shall describe the growth of citations between publications.
There, we make use of five parameters \(\vecb{\theta}=[\alpha,\beta,\gamma,\delta,\tau]\) that will allow us to include or exclude different mechanisms that impact citations.
Their importance can be tested in a statistical approach described in the following section.

\section{Time-respecting Maximum Likelihood Estimation}
\label{sec:temp-mle}

\subsection{Estimation of model parameters}\label{sec:estim-model-param}

Having defined multi-layer network growth by means of model components \(\pi^{\mu\nu}_{ij}\left[G(t);\vecb{\theta}\right]\), here we describe a maximum likelihood estimation of the model parameter vector \(\vecb{\theta}\), given a growing network \(G(t)\).
Similar to the approach in~\cite{Medo2014,Leskovec2008}, we calculate the model likelihood under the assumption of statistical independence of \emph{adding} individual edges:
\begin{equation}
  \ln \mathcal{L}\left[\vecb{\theta}; G(T)\right] = \sum_{t=1}^{T-1} \sum_{\mu,\nu} \sum_{\substack{e_{ij} \in E^{\mu\nu}(t+1)\\ e_{ij} \notin E^{\mu\nu}(t)}} \ln \pi^{\mu\nu}_{ij}\left[G(t);\vecb{\theta}\right],
  \label{eq:mle:ll}
\end{equation}
where \(T\) is the last observed time.
Note that the independence assumption does not apply to the edges themselves: the probability of an edge depends on the network state at time $t$ and thus on the previously existing edges.
But adding one edge at a given time does not depend on other edges added at the same time.
The vector of model parameters \(\vecb{\hat\theta}\) that maximises the likelihood of the model given the network is found by solving the equation
\begin{equation}
  \label{eq:mle:score}
  \left. \frac{\partial \ln \mathcal{L}}{\partial \vecb{\theta}} \right|_{\vecb{\hat\theta}} = 0.
\end{equation}
Equation \eqref{eq:mle:score} allows us to find the best estimate of the given model's parameter vector. However, it does not provide information about the \emph{uncertainty} of the estimates.
Fortunately, the latter can also be computed as part of the Maximum  Likelihood  Estimation  (MLE) method.   It  is  an  important  property  of  the  MLE  method that, under some mild conditions, the vector of parameter estimates \(\mathbf{\hat\theta}\) follows a multivariate normal distribution around the true value of the parameters.
This means that, the uncertainty about the parameter estimates can be adequately described using standard errors based on the corresponding covariance matrix.
In the context of MLE, the covariance matrix of the parameters is then given by the inverse of the \emph{Fisher information matrix} \(\mathcal{I}(\mathbf{\theta})\):
\begin{equation}
  \sigma^2(\vecb{\hat\theta}) = {[\mathcal{I}(\vecb{\hat\theta})]}^{-1}.
  \label{eq:mle:var}
\end{equation}
The matrix \(\mathcal{I}(\vecb{\hat\theta})\) is the expectation of the second derivatives of the log-likelihood function, which is usually not feasible to find.
Hence, in practice it is commonly approximated by the \emph{observed} Fisher information matrix:
\begin{equation}
  \mathcal{I}(\vecb{\theta}) \approx \mathcal{I}(\vecb{\hat\theta}) = - \left. \frac{\partial^2 \ln \mathcal{L} (\vecb{\theta})}{\partial \vecb{\theta} \, \partial \vecb{\theta}} \right|_{\vecb{\hat\theta}}.
    \label{eq:mle:fisher}
\end{equation}
I.e., $ \mathcal{I}(\vecb{\theta})$ corresponds to the values of the second derivatives as in \cref{eq:mle:fisher} calculated for the parameter estimate \(\vecb{\hat\theta}\).
In general, the \gls{mle} of the parameters \(\vecb{\theta}\) is not performed by finding the maxima of \cref{eq:mle:score} analytically  but is computed by means of numerical optimisation as discussed below in \cref{sec:computation}.

\subsection{Model selection}\label{sec:model-selection}

So far, we have specified our growth model by means of model components \(\pi^{\mu\nu}_{ij}\left[G(t);\vecb{\theta}\right]\),
where the  model parameter vector \(\vecb{\theta}\) was determined by a MLE.
Model selection denotes a defined procedure of statistically comparing different models in their ability to capture the data given with the growing network $G(t)$.
With ``different models'', we not only refer to different parameter vectors \(\vecb{\theta}\), but also to different formulations of sets of model components \(\pi_{ij}^{\mu\nu}\).
For instance, we can define a model $A$ that implements information from two sets of edges, \(E^{\mu\nu}\) and \(E^{\mu'\nu'}\), and then compare it to a model $B$ based on only one set of edges.
In general, model $A$ is expected to describe the data better, in statistical terms, because it takes more data into account.
Therefore, model selection has to consider not only the goodness of a fit, but also the \emph{model complexity} as expressed by the degrees of freedom of the model: the more information we put into the model, the more complex it is.

Later, in \cref{sec:scientometric-results}, we will define seven models for the growth of citations among scientific publications.
In some of these models the new citations will depend only on the previous citations, while in other models the new citations will also depend on authorship relations. To compare these models, we will compute their \emph{relative likelihoods}, similar to~\cite{Medo2014}:
\begin{equation}
  \label{eq:mle:rel-lik}
  w_m = \exp \left\{\frac{1}{2} \left[\min_{m}(AIC_m) - AIC_m \right] \right\} 
\end{equation}
where \(AIC_m\) is the value of Akaike Information Criterion of the model \(m\).
It is based on the maximum likelihood value of the model, corrected for the model complexity:
\begin{equation}
  \label{eq:aic}
  AIC_m = - 2 \log \mathcal{L}[\vecb{\hat\theta}_m; G(T) ] + 2|\vecb{\theta}_m|,
  \end{equation}
where \(|\vecb{\theta}_m|\) denotes the size of vector \(\vecb{\theta}_m\), i.e., the number of parameters of the model \(m\).
For better interpretability, after computing the values of \(w_m\) according to \cref{eq:mle:rel-lik}, they are normalised to add to one.
Our choice of the relative likelihoods \(w_m\) of models is motivated by their applicability for comparing a wide range of models.
This is in contrast with, e.g., likelihood ratio tests that could be used only for comparing so-called nested models where one model is a special case of another.

\subsection{Scalable \gls{mle} by event sampling}\label{sec:computation}
While the MLE cannot be solved \emph{analytically}, we can make use of the fact that, for commonly studied network growth models, the likelihood function has a convex shape (see \cref{sec:app} and \cite{Medo2014}).
This means that the \gls{mle} can be done \emph{numerically} by using a greedy hill-climbing algorithm.
In the following, we will use the Broyden-Fletcher-Goldfarb-Shanno (BFGS) algorithm.
It has the advantage that, while finding the function extremum, it also estimates the inverse of the Hessian matrix, which for the log-likelihood function corresponds to the Fisher information matrix.
This means that we obtain both the MLE of the parameters and their variances at the same time (see \cref{eq:mle:fisher}).
More specifically, to obtain our results we use a variant of the algorithm, \texttt{L-BFGS-B}.
It allows to set boundary constraints on the parameters and uses limited memory, making the whole procedure more scalable~\cite{Byrd1995,Zhu2997}.
\texttt{L-BFGS-B} is a standard algorithm for solving nonlinear optimisation problems with bounded variables and implementations are available in common scientific computing libraries \cite{scipy}.
For further details on the algorithm's complexity, we refer to the original publications \cite{Byrd1995,Zhu2997}.

Performing a MLE for  \cref{eq:mle:ll} based on the full network data becomes very costly for large networks.
For each time step the computations for \(\pi^{\mu\nu}_{ij}[G(t);\vecb{\theta}]\) have to be redone based on the network state at that time.
Therefore, the total computation time scales at least with the square of the final network size.
The larger the network, the more distinct time steps we have to consider to describe its growth (as time is  parametrised by edges added, see \cref{sec:model-general}).
To reduce the computational cost of the MLE, we apply the model evaluation of \cref{eq:mle:ll} only to a \emph{subset} of growth events observed over time \(t \in [1,T]\).
For this, we sample, without replacement, a number \(S < T\) of time steps uniformly at random, \(\{t_s\}_{s \in [1,S]}\).
This set of time stamps uniformly covers the full time range \([1,T]\) of the network growth, provided that  the largest sampled time step \(t_S\) is close to the final time \(T\).
Then, we can compute the MLE of a model based on this sample of time steps:
\begin{equation}
  \ln \mathcal{L}\left(\vecb\theta; G(T), S\right) = \sum_{s=1}^{S} \sum_{\mu,\nu} \sum_{\substack{e_{ij} \in E^{\mu\nu}(t_s+1)\\ e_{ij} \notin E^{\mu\nu}(t_s)}} \ln \pi^{\mu\nu}_{ij}[G(t_s);\vecb{\theta}].
  \label{eq:mle:ll:sample}
\end{equation}
Through the model components \(\pi^{\mu\nu}_{ij}[G(t_s);\vecb{\theta}]\) 
potentially all nodes and edges present in \(G(T)\) will affect the log-likelihood of the model given by \cref{eq:mle:ll:sample}. 
In other words, the likelihood estimation is based on `probes' of temporal events over the whole growth period of the network.
By fixing the number of time steps, \(S\), the computation time can, for the final network size,
be decreased by an order of magnitude. 
As a  trade-off, the precision of the parameter estimates will be lower because their variance is inversely proportional to the number of observations used in the estimation.
This trade-off can be controlled by adjusting the number \(S\) of sampled time stamps according to the desired precision of the parameter estimate.

\subsection{Temporal stability of a model}
\label{sec:trends}

The above discussion assumes that the growth model does not \emph{explicitly} depend on time, even though the studied network changes over time.
In particular, the parameter vector \(\vecb{\theta}\) used to define the model components \(\pi^{\mu\nu}_{ij}[G(t_s);\vecb{\theta}]\) does not depend on time.
To confirm that this assumption holds we split the whole growth time period \([1,T]\) into 
into \(R\) consecutive time windows such that each of them contains approximately \(\lfloor T/R \rfloor\) time steps. 
We then partition the log-likelihood function, \cref{eq:mle:ll}, into the sum of log-likelihoods computed for each time window separately:
\begin{equation}
	\ln \mathcal{L}\left[\vecb{\theta}; G(T)\right] = \sum_{r=1}^{R} \ln \mathcal{L}\left[\vecb{\theta}; G, r\right],
  \label{eq:mle:partition}
\end{equation}
where each summand on the right hand side is given by
\begin{equation}
	\ln \mathcal{L}\left[\vecb{\theta}; G, r\right] = \sum_{t=(r-1)\lfloor \frac{T}{R} \rfloor}^{r\cdot\lfloor \frac{T}{R} \rfloor}
	\sum_{\mu,\nu} \sum_{\substack{e_{ij} \in E^{\mu\nu}(t+1)\\ e_{ij} \notin E^{\mu\nu}(t)}} \ln \pi^{\mu\nu}_{ij}\left[G(t);\vecb{\theta}\right].
  \label{eq:mle:window}
\end{equation}
\cref{eq:mle:window} provides the log-likelihood of the model computed for a certain time window \(r \in [1,R]\).
Based on this, we can estimate the model parameters for each time window separately using the MLE described earlier. 
The assumption of a stationary model that itself does not depend on time is tested by comparing the parameter estimates for the \(R\) consecutive time windows.
We will use this approach in the next section to validate the method based on synthetically generated networks with known growth mechanisms.

\section{Method Validation on Synthetic Data}\label{sec:valid-synth-data}

To confirm that the presented method leads to correct results,
we first use a synthetic network comprising two sets of nodes denoted as \(V^{p}\) and \(V^{a}\) and two sets of directed edges \(E^{pp}\) and \(E^{ap}\).
That is, we have a two-layer network with edges within layer $p$ and between the two layers, $p$ and $a$.
The set $V^{a}$ consists of a fixed number of initially disconnected nodes.
During the growth of the network, these nodes will be subsequently connected to the nodes in layer $p$, as described below. 

Hence, for the growth of the synthetic network we 
concentrate on layer $p$, where one node \(i\) at a time is added with an \emph{out-degree} randomly drawn between one and five. 
This node connects to already existing nodes
based on  \emph{linear preferential attachment}, i.e. on the \emph{in-degrees} of the chosen nodes.
Because of the two-layer network, nodes are connected \emph{within} and \emph{across} layers.
Consequently, the first growth mechanism takes the in-degree of nodes \emph{within} a layer, i.e., the edges in the set \(E^{pp}\),  into account.
The second growth mechanism, on the other hand,  builds on the in-degree of nodes \emph{across} layers, i.e., the  edges in the set \(E^{ap}\). 

Using the combined information from the two layers, we specify the growth mechanism  as follows: when adding an edge from the newly added node \(i \in V^p\), we choose the existing node \(j \in V^p\) with probability: 
\begin{align}
  \label{eq:pi11}
  \pi^{pp,\mathrm{true}}_{ij}[G(t);\alpha,\delta] =&\alpha \frac{k^{pp,in}_j(t) + \delta}{\sum_{l \in V^p(t)} [k^{pp,in}_l(t) + \delta]} + (1-\alpha) \frac{k^{ap}_j}{\sum_{l \in V^p(t)} k^{ap}_l} ,
\end{align}
where \(k^{pp,in}_j(t)\) is the number of incoming edges of node \(j \in V^p\) within layer $p$, i.e., its in-degree with respect to the edge set \(E^{pp}(t)\), and \(k^{ap}_j(t)\) is the number of edges of node \(j\) to nodes in layer $a$,  i.e., its in-degree with respect to the edge set \(E^{ap}(t)\). 
The parameter \(\alpha \in [0,1]\) weights between the two growth mechanisms (often called mixture weight).

The constant \(\delta\) \cite{Price1976,Barabasi1999}, which appears in the first growth mechanism, is added to the in-degree of nodes in layer $p$ to ensure that, in the case of directed networks, a newly added node with non-zero out-degree, but \emph{zero in-degree} can attract incoming edges.
For the second mechanism we do not need to include such an additive constant because each node in \(i \in V^p\) has at least one edge \(e_{ij} \in E^{ap}\) connecting it to a node \(j \in V^a\) in the second layer.
Specifically,
we connect the newly added node \(i \in V^p(t)\) to between one and five nodes in layer $a$ with a uniform probability:
\begin{equation}
  \label{eq:pi21}
  \pi^{ap,\mathrm{true}}_{ji}[G(t)] = \frac{1}{|V^a(t)|}.
\end{equation}
For the experiment below, we choose \(V^p(T) = V^a(T) = 2000\) nodes in each of the two layers.
Adding one node at a time to layer $p$, this implies a total growth period of \(T=2000\).
To test whether our procedure works, we fix the values of the parameters as \(\delta = 1.2\) and \(\alpha = 0.7\). 
Once we generate a synthetic network, we can apply the MLE to recover the (known) parameters of the model.
Given the sampling procedure described in Sect.~\ref{sec:computation}, we estimate the parameters for varying sizes \(S \leq T\) of the randomly sampled time stamps. 
Additionally, we repeat the procedure ten times for each sample size.

\begin{figure}[htbp]
  \centering
  \includegraphics[width=.35\textwidth]{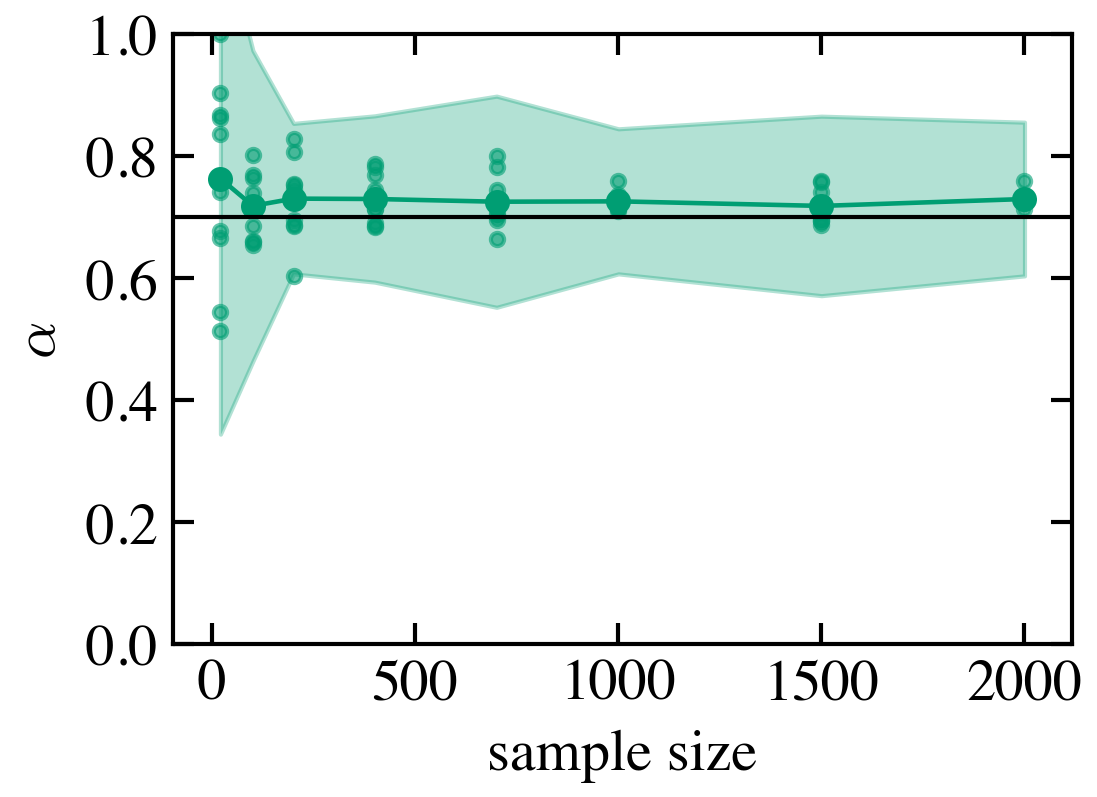} 
  \includegraphics[width=.35\textwidth]{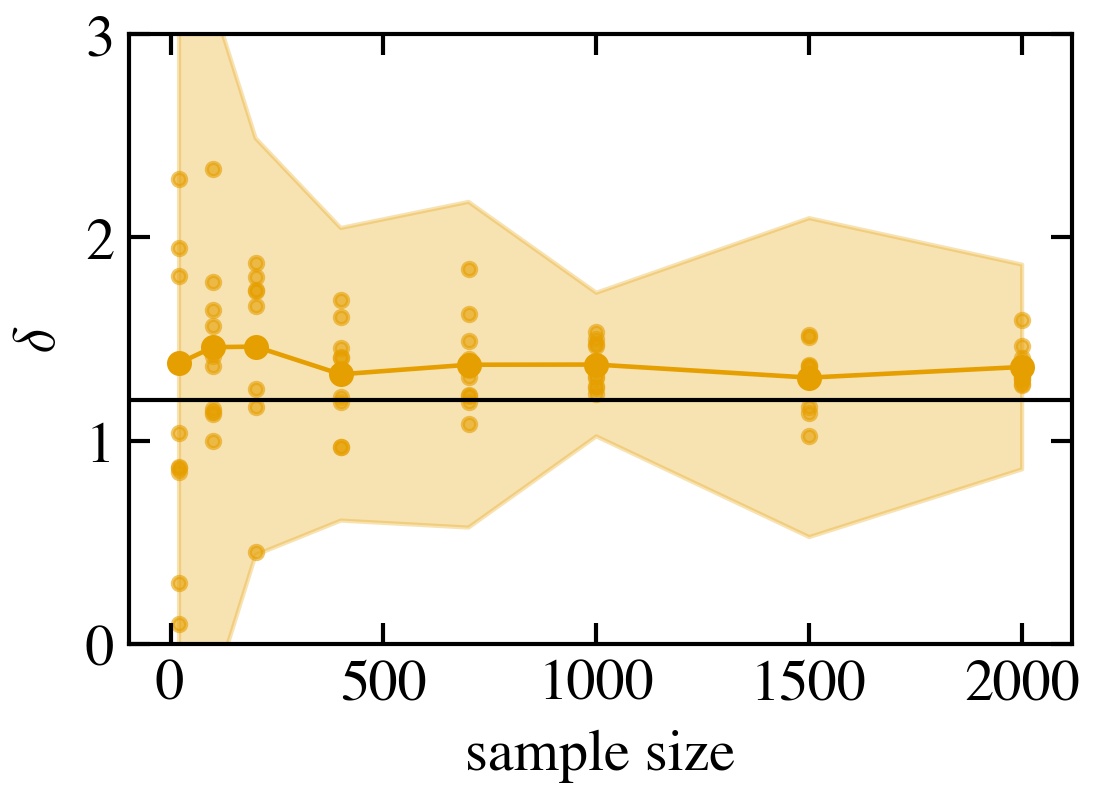}
  \caption{Parameter estimates (small dots) of the model given in \cref{eq:pi11} for varying sample size, with \(10\) realisations for each sample size.
  These estimates (large dots and lines) and their standard errors (shaded area) are pooled for each sample size.
  	The black line corresponds to the true parameter value used for generating the networks.}\label{fig:param-size}
\end{figure}
The results of this numerical experiment are reported in \cref{fig:param-size}.
We find that (i) the estimates are close to the true value and the true value falls within one standard error of the estimates, (ii) the variance decreases with the sample size as expected.
The decrease becomes noticable beyond a  sample size of about 400, i.e. 20\% from the full data.
The results confirm that our method is able to identify the correct parameters. 

\begin{figure}[htbp]
  \centering
  \includegraphics[width=.35\textwidth]{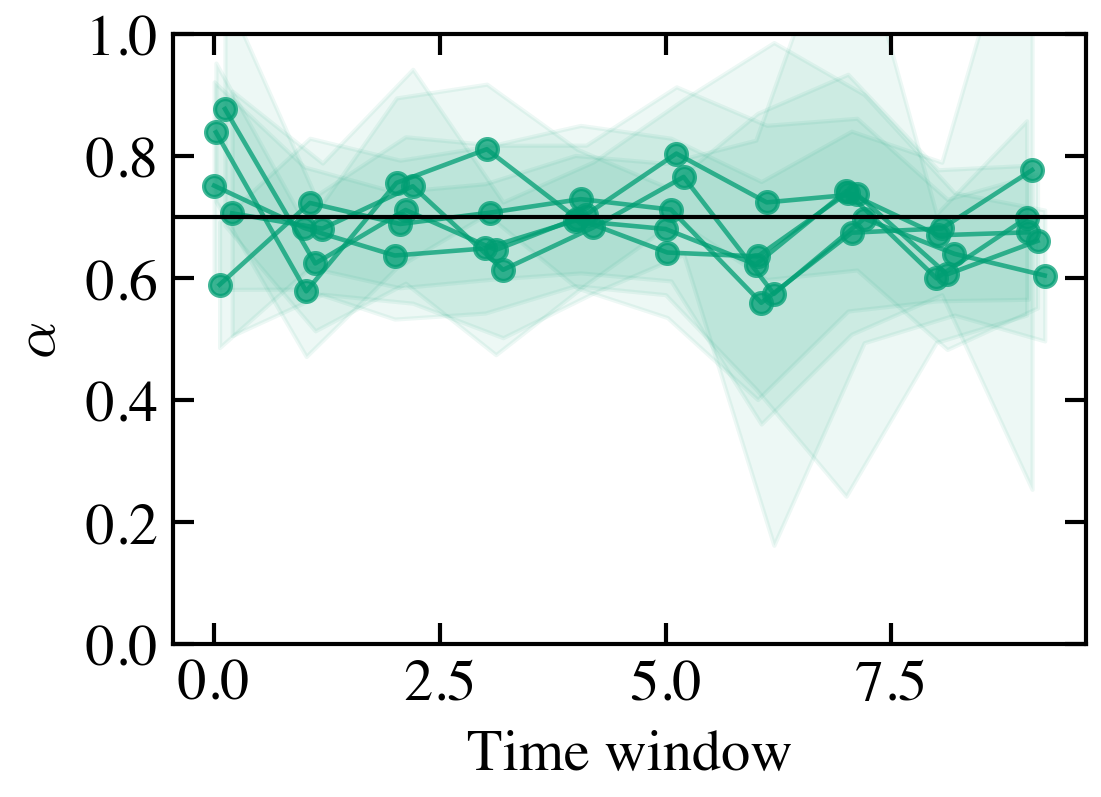} 
  \includegraphics[width=.35\textwidth]{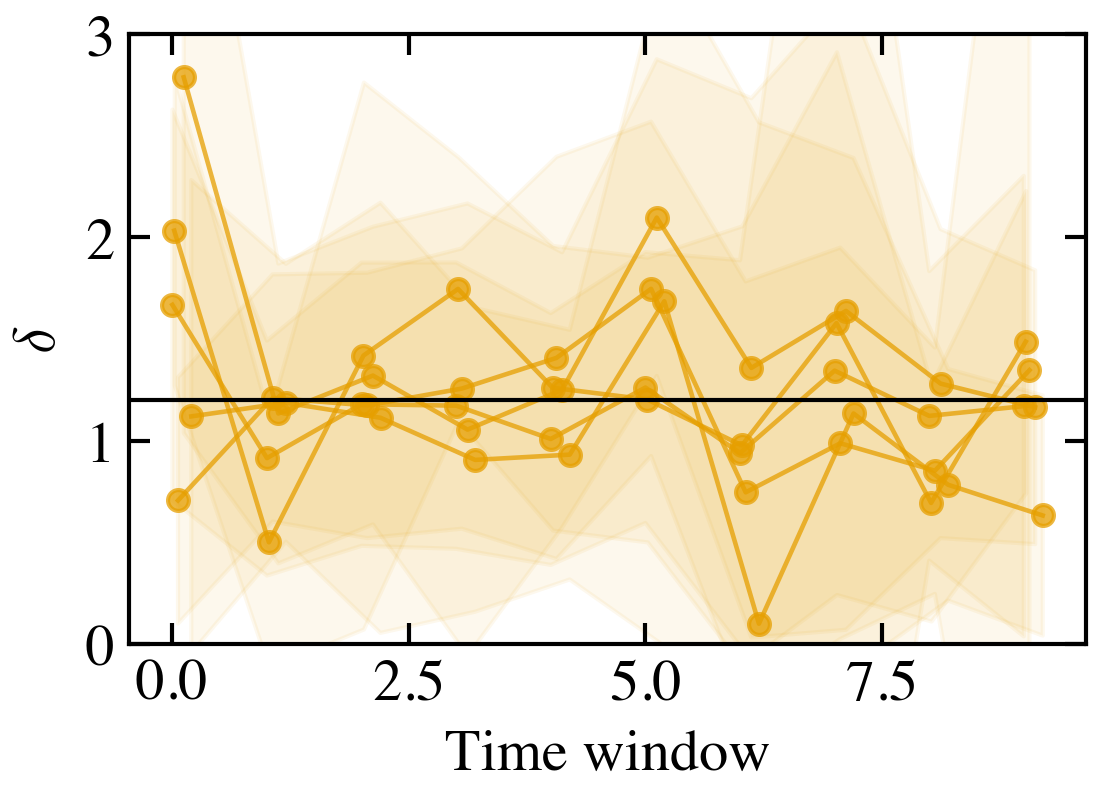}
  \caption{The observed growth of a network is partitioned into ten consecutive time windows, each comprising approximately the same number of time steps.
	The model in \cref{eq:pi11} is then evaluated based on each time window separately.
        The resulting parameter estimates (dots and lines) and their standard errors (shaded areas) are shown for five synthetic networks. The black line corresponds to the true parameter value.
      }\label{fig:param-trend}
\end{figure}
Next, we want to test whether the parameter estimates are stable over different growth periods of the network, as described in \cref{sec:trends}.
For this test, we partition the total growth period of \(T=2000\) time stamps
into \(R=10\) time windows.
We then estimate the model parameters for each time window based on the log-likelihood defined in \cref{eq:mle:window}.
The results for five synthetically generated networks are shown in \cref{fig:param-trend}.
We see that for both parameters the true value lies within one standard error of the estimates for each time window.
Further, no trend is observed in the parameter estimates over time.
This confirms that our method is robust also with respect to the sampling procedure. 

Last but not least, it is important to also demonstrate that the \emph{model selection approach} based on the relative likelihoods (\cref{sec:model-selection}) can correctly identify the best model among various candidates.
To this end, we eventually evaluated a set of different models  (explained in the next section) in addition to the known true model of \cref{eq:pi11}.
We show these results in the Appendix \cref{tab:growth:fits-sinthetic}.
They confirm that, indeed, the model selected this way corresponds to the true model used to generate the synthetic data set, i.e. that the correct model was identified.

\section{Model Specification for Citation Dynamics}
\label{sec:model-spec-citat}

Now that we have seen how our modeling framework can be used and how it is able to correctly identify the mechanisms underlying a synthetic growth process, we are ready to specify the framework for the real-world application in scientometrics.
We recall that model specification implies to determine the model components $\pi^{\mu\nu}_{ij}\left[G(t);\vecb{\theta}\right]$ from Eq.~(\ref{eq:general-model}). 
Specifically, to model the two-layer network with publications $V^p$ and authors $V^a$ we need to determine the four model components shown in Figure \ref{fig:multi-layer}.
This requires us to come up with some dynamic hypotheses of how these two layers are coupled, which should be aligned to the empirical facts.

\subsection{The model component $\pi_{ij}^{pp}$}
\label{sec:model-component}

Given the dynamics for our network illustrated in Figure \ref{fig:net-evol}, we can make two observations.
First, authors do not have direct links, but only via co-authored publications.
Therefore, the probability $\pi^{aa}_{ij}[\cdot]$ does not need to be modelled.
Second, relations between authors and papers are undirected, and thus $\pi^{ap}_{ij}[\cdot] = \pi^{pa}_{ij}[\cdot]$ holds.
Furthermore, we assume that the links between authors and publications are given by the data, making $\pi^{ap}_{ij}[\cdot]$ deterministic.
Therefore, we are only left with specifying a single model component, $\pi^{pp}_{ij}[\cdot]$.
This defines the probability of an incoming publication $i$ to form an edge, i.e., to cite, an existing publication $j$.
In most general terms, this probability can be written in full as:
\begin{align}
  \label{eq:model_component}
  \pi_{ij}^{pp}[G(t); \vecb{\theta}] &= \alpha\dfrac{\eta_j\left[G(t)\right]^\beta \cdot K_j\left[G(t); \gamma, \delta, \tau\right]}{\sum\limits_{l \in V^p(t)}\eta_l\left[G(t)\right]^\beta  \cdot K_l\left[G(t); \gamma, \delta, \tau\right]} + (1-\alpha)\dfrac{\eta_j\left[G(t)\right]}{\sum\limits_{l\in V^p(t)}\eta_l\left[G(t)\right]},\\
  \vecb{\theta} &= \begin{bmatrix}\alpha, & \beta, & \gamma, & \delta, & \tau \end{bmatrix}^{'}
\end{align}
which follows the structure already discussed for Eq. (\ref{eq:pi11}). 
Note that once published, the citations in publication are generally not further modified.
Hence, we do not model the probability of existing publications to form new links.
Equation~(\ref{eq:model_component}) comprises two constituents, $K_j[\cdot]$ and $\eta_j[\cdot]$:
With $K_j[\cdot]$, we specify the \emph{citation self-dynamics}, in particular different specifications from the preferential attachment family.
$\eta_j[\cdot]$ shall consider the \emph{fitness} of the publication.
It describes, for instance, how attractive the publication is for citations~\cite{Wang2013} and relates to the fitness term in other preferential attachment models~\cite{Bianconi2001,Pham2016}.
We want $\eta_j[\cdot]$ to capture \emph{non-citation related influences}, specifically influences from the authors of the paper on its probability to receive citations.
Therefore, we will refer to $\eta_j[\cdot]$ as the \emph{social constituent}.
Both components will be individually addressed in the following sections.

The parameters $\alpha \in [0,1]$ and $\beta$ can be used to couple the citation and the social constituents either additively or multiplicatively.
For $\beta = 0$, we obtain
\begin{align}\label{eq:model_component_additive}
\pi_{ij}^{pp}[G(t); \vecb{\theta}] &= \alpha\dfrac{K_j\left[G(t); \gamma, \delta, \tau\right]}{\sum\limits_{l \in V^p(t)} K_l\left[G(t); \gamma, \delta, \tau\right])} + (1-\alpha)\dfrac{\eta_j\left[G(t)\right]}{\sum\limits_{l\in V^p(t)}\eta_l\left[G(t)\right]},
\end{align}
which means the probability $\pi_{ij}^{pp}[\cdot]$ for publication $i$ to cite publication $j$ is the weighted mixture of the two constituents.
Such \emph{mixture models} are used when the statistical population is known to comprise sub-populations that are described by different probability distributions.
The mixture weight $\alpha$ is, then, determined by the relative size of each sub-population.
In our case, it means that node $j$ is chosen \emph{either} based on constituent \(K_j[\cdot]\) of citation self-dynamics, \emph{or} based on the social constituent \(\eta_j[\cdot]\), with the respective weight $\alpha$.

For $\alpha = 1$, we obtain multiplicative coupling:
\begin{align}
  \label{eq:1}
\pi_{ij}^{pp}[G(t); \vecb{\theta}] = \dfrac{\eta_j\left[G(t)\right]^\beta \cdot K_j\left[G(t); \gamma, \delta, \tau\right]}{\sum\limits_{l \in V^p(t)}\eta_l\left[G(t)\right]^\beta  \cdot K_l\left[G(t); \gamma, \delta, \tau\right]}
\end{align}
It means that the social constituent \emph{scales} the effect of the citation constituent during the growth process, with 
$\beta \in [0, \infty)$ as the \emph{impact}  of the social constituent. 
Hence, the growth mechanism defined within the citation layer is the main, \emph{baseline}, effect while the social constituent biases it.
For intermediate values of $\alpha$ and $\beta$ mixtures between additive and multiplicative coupling can be achieved.

\subsection{Citation Constituent}\label{sec:growth:cit-comp}

To define the citation constituent $K_j[\cdot]$, we use a general expression for preferential attachment:
\begin{align}\label{eq:2}
	K_j\left[G(t); \vecb{\theta}\right] =[k_j^{pp,\text{in}}(t) + \delta]^\gamma \cdot e^{-\frac{t-t_j}{\tau}}
\end{align}
The term $k_j^{pp,\text{in}}(t) + \delta$ describes linear preferential attachment, where $k_j^{pp,\text{in}}(t)$ denotes the in-degree of the publication $j$ in the citation layer at time $t$, i.e., before citations made to $j$ are considered.
The constant $\delta$, again, corrects for publications without citations (i.e. in-degree zero) at time $t$~\cite{Sen2005}.
The parameter $\gamma \in [0, \infty)$ allows us to study \emph{non-linear preferential attachment} by controlling the degree-related preference.
Specifically, it amplifies or weakens the effect of the number of previous citations.
Finally, the term $e^{-\nicefrac{(t-t_j)}{\tau}}$ introduces a \emph{relevance decay}~\cite{Medo2011}.
Here, $t_j$ denotes the time at which the publication $j$ was added to the network, thus the probability for a publication to be cited now also depends on its age.
For simplicity, we will measure time using the number of publications in layer $p$.
This means \(t_j = j\) if nodes are labeled in the order they were added to the network.
The parameter $\tau \in [0, \infty)$ determines the characteristic time of the relevance decay.

In our empirical analysis, we study four specifications of the citation constituent \(K_j[\cdot]\) defined by certain settings of the parameters $\gamma$, $\delta$, and $\tau$ as follows: 
\begin{description}
        \item[Uniform Attachment (UNIF)]
This model defines a baseline without citation preferences for any paper. This is achieved, e.g. with $\gamma = \delta = 0$ and $\tau = \infty$, yielding
\begin{align}
  \label{eq:3}
K_j\left[G(t); \vecb{\theta}\right] = 1.
\end{align}

\item[Preferential Attachment (PA)] 
This model is motivated by the general finding that highly cited publications tend to receive more citation in the future~\cite{Price1976}.
This can be described by linear preferential attachment, which is obtained for $\gamma = 1$ and $\tau = \infty$.
\begin{align}
K_j\left[G(t); \vecb{\theta}\right] = k_j^{pp,\text{in}}(t) + \delta
\label{eq:PA}
\end{align}

\item[Preferential Attachment with Relevance Decay (PA-RD)]
  The motivation for this model is similar to \texttt{PA}, but it additionally considers that papers are cited less over time~\cite{Medo2011}, for which the most evidence can be found in literature.
This is obtained for $\gamma = 1$.
\begin{align}
  \label{eq:4}
K_j\left[G(t); \vecb{\theta}\right] = [k_j^{pp,\text{in}}(t) + \delta] \cdot e^{-\frac{t-t_j}{\tau}}
\end{align}

\item[Non-Linear Preferential Attachment with Relevance Decay (PA-NL-RD)]
The added non-linearity to the \texttt{PA-RD} model allows to scale the citation probability with the current citation count~\cite{Sen2005,Golosovsky2013}. 
To keep the same number of degrees of freedom for this model compared to PA-RD, we fix the offset $\delta=1$ yielding
\begin{align}
  \label{eq:5}
K_j\left[G(t); \vecb{\theta}\right] = [k_j^{pp,\text{in}}(t) + 1]^\gamma \cdot e^{-\frac{t-t_j}{\tau}}.
\end{align}

\end{description}

\subsection{Social Constituent}\label{sec:growth:soc-comp}

To completely specify the model component $\pi_{ij}^{pp}[\cdot]$, Eq.~(\ref{eq:model_component}), we have to make assumptions about the 
social constituent, \(\eta_j\left[G(t)\right]\).
Here we mostly follow the arguments made by \citeauthor{Merton1968} in his seminal paper~\cite{Merton1968}, where he theorizes that the academic prominence of authors influences the success and recognition of their future work.
To quantify different levels of prominence, we use information related to the number of co-authors and the number of their
previous publications, as specified in the following:

\begin{description}

\item[Number of Authors (NAUT)]
As the simplest possibility, we define the social constituent as the number of authors that wrote the publication.
We argue that more co-authors (i) increase a paper's potential social exposure, (ii) allow for a better division of labour and to a higher impact~\cite{Leimu2005}.
A paper with a higher number of co-authors should therefore receive more citations. Mathematically, this is formulated as
\begin{equation}
	\eta_j\left[G(t)\right] = k^{ap}_j(t),
	\label{eq:coupling:soc:naut}
\end{equation}
where $k^{ap}_j(t)$ counts the number of authors of a publication.

\item[Number of Previous Co-authors (NCOAUT)]
  We  assume that authors who have previously collaborated with a larger number of co-authors have more experience and more visibility in their community, which in turn increases the citations for their new publications. 
To calculate, for the co-authors of a given paper, their set of unique \emph{previous} co-authors, we use: 
\begin{equation}
\eta_j\left[G(t)\right] = \big| \{ v \given \exists \lambda^{ap}_{3,jv}(t) \} \big|,
\label{eq:coupling:soc:ncoaut}
\end{equation}
That means, we count the unique endpoints \(v\) of the self-avoiding paths \(\lambda^{ap}_{3,jv}\) of length three over author-publication edges, i.e., paths \texttt{publication $\rightarrow$ authors $\rightarrow$ previous publications $\rightarrow$ previous co-authors}.

\item[Maximum Number of Previous Co-authors (MAXCOAUT)]
  In this variant of  \texttt{NCOAUT} we only take the author with the largest number of previous co-authors into account~\cite{Sarigol2014}.
This is motivated by \citet{Merton1968}, who argues that the academic credit for a publication written by a team of authors is often attributed to the most prominent author. For authors \(q \in V^a(t)\) of the publication \(j\) this yields
\begin{equation}
\label{eq:coupling:soc:maxcoaut}
\eta_j(t) = \max_q ( \big|\{ v  \given \exists \lambda^{ap}_{2,qv}(t) \} \big|)
\end{equation}
That means, we count the unique endpoints \(v\) of the self-avoiding paths \(\lambda^{ap}_{2,jv}\) of length two: \texttt{author $\rightarrow$ previous publications $\rightarrow$ previous co-authors}.

\item[Number of Previous Publications (NPUB)]
We argue that with more prior publications, i.e., with increasing experience of an author, (i) the impact of new publications and  (ii) the author's recognition in the academic community should both increase, which leads to increased citation counts of future publications.
Hence, we consider the number of distinct publications written by all authors of  publication \(j\).
\begin{equation}
\eta_j(t) = \big| \{ v \given \exists \lambda^{ap}_{2,jv}(t) \} \big|
\label{eq:coupling:soc:npub}
\end{equation}
Similar to \cref{eq:coupling:soc:ncoaut,eq:coupling:soc:maxcoaut}, we count the unique endpoints \(v\) of the self-avoiding paths \(\lambda^{ap}_{2,jv}\) of length two: \texttt{publication $\rightarrow$ authors $\rightarrow$ other publications}.

\item[Maximum Number of Previous Publications (MAXPUB)]
  In this variant of \texttt{NPUB} we assume that the probability to be cited only depends on the maximum number of previous publications among all co-authors.
For authors \(q \in V^a(t)\) of the publication \(j\), this is formulated as
\begin{equation}
\eta_j(t) = \max_q ( k^{ap}_{q}(t) ),
\label{eq:coupling:soc:maxpub}
\end{equation}
where $k^{ap}_{q}$ is the degree of author $q$ with respect to author-publication edges at time $t$.
\end{description}

Based on the three alternatives for the citation constituent and the five alternatives for the social constituent, we can now create a large number of possible models, i.e. expressions for $\pi_{ij}^{pp}[\cdot]$, Eq.~(\ref{eq:model_component}), by using either additive or multiplicative coupling of the constituents.
Three models (\texttt{UNIF}, \texttt{PA-RD} and \texttt{PA-NL-RD}) consider only the citation self-dynamics, i.e., there is no coupling to the social constituent. 
Because we take from the literature that \texttt{PA-RD} is the most promising candidate for the citation constituent, we restrict the combined models only to combinations of \texttt{PA-RD}.
For the additive combination, we test it together with all five alternatives for the social constituent, for the multiplicative combination only with the two most promising ones. 
A summary of all tested models and their free parameters is given in Table \ref{tab:model_specification}.

\begin{table*}[b!]
  \caption{Specification and relationships of the seven models used in the analysis. Columns with \faQuestion~denote the free parameters that are fitted for the model.
    (PA): plain preferential attachment, 
    (PA-RD): preferential attachment with relevance decay,
    (PA-NL-RD): preferential attachment with with relevance decay and added non-linearity. In addition we consider models that combine PA-RD
    with different candidates for the social constituent (see Sect.~\ref{sec:growth:soc-comp} for explanation). All models are compared to a baseline model with uniform attachment (UNIF).
  }
	\label{tab:model_specification}
	\begin{tabularx}{\textwidth}{lf{3cm}f{3cm}CCCCC}
		\toprule
		\textbf{Model} & \makecell{\textbf{Citation}\\ \textbf{Constituent}} & \makecell{\textbf{Social}\\ \textbf{Constituent}} & $\alpha$ &  $\beta$ & $\gamma$ & $\delta$ & $\tau$\\\midrule
		\textbf{UNIF} & UNIF & --- & 1 & 0 & 0 & 0 & $\infty$ \\
		\textbf{PA} & PA & --- & 1 & 0 & 1 & \faQuestion & $\infty$ \\
		\textbf{PA-RD} & PA-RD & --- & 1 & 0 & 1 & \faQuestion & \faQuestion \\
		\textbf{PA-NL-RD} & PA-NL-RD & --- & 1 & 0 & \faQuestion & 1 & \faQuestion \\\cmidrule(lr){1-8}
		\textbf{PA-RD + NAUT} & PA-RD & NAUT & \faQuestion & 0 & 1 & \faQuestion & \faQuestion \\
		\textbf{PA-RD + NCOAUT} & PA-RD & NCOAUT & \faQuestion & 0 & 1 & \faQuestion & \faQuestion  \\
		\textbf{PA-RD + MAXCOAUT} & PA-RD & MAXCOAUT & \faQuestion & 0 & 1 & \faQuestion & \faQuestion  \\
		\textbf{PA-RD + NPUB} & PA-RD & NPUB & \faQuestion & 0 & 1 & \faQuestion & \faQuestion  \\
		\textbf{PA-RD + MAXPUB} & PA-RD & MAXPUB & \faQuestion & 0 & 1 & \faQuestion & \faQuestion  \\\cmidrule(lr){1-8}
		\textbf{PA-RD $\times$ NCOAUT} & PA-RD & NCOAUTH & 1  & \faQuestion & 1 & \faQuestion & \faQuestion \\
		\textbf{PA-RD $\times$ NPUB} & PA-RD & NPUB & 1  & \faQuestion & 1 & \faQuestion & \faQuestion \\
		\bottomrule
	\end{tabularx}
\end{table*}

\section{Empirical Authorship-Citation Networks}
\label{sec:scientometric-results}

The bibliographic data used for our model testing comes from two sources: APS journals~\footnote{Data provided by American Physical Society for research purposes, see \protect\url{https://journals.aps.org/datasets}} and INSPIRE~\footnote{See \protect\url{http://inspirehep.net/info/general/project/index} for general information and \protect\url{http://inspirehep.net/dumps/inspire-dump.html} for data acquisition}.
The details of these data, their origin and our motivation to use them, are described in the \cref{sec:appendix_data}.
Here, we only list the nine different journals that were included in our study:
\begin{itemize}[noitemsep, topsep=0pt]
\item \gls{jpr}
\item \gls{jpra}
\item \gls{jprc}
\item \gls{jpre}
\item \gls{jrmp}
\item \gls{jhep}
\item \gls{jpr_ins}
\item \gls{jpl}
\item \gls{jnp}
\end{itemize}

\subsection{Results for nine empirical citation networks}\label{sec:results:empirical_networks}

In the following, we fit, for each of the nine citation networks listed above, our eleven model specifications for \(\pi_{ij}^{pp}[G(t); \vecb{\theta}]\) presented in \cref{tab:model_specification}. 
Subsequently, we apply the model selection technique described in \cref{sec:temp-mle} to select the most adequate model.

One could perform a step wise model selection as follows:
For each two-layer network, first fit the models with only the citation constituent and choose the one with the highest relative likelihood.
Then, consider models with couplings between citation and social constituents based on the previously selected citation constituent.
Instead, we chose the fixed set of models with and without coupling based on our prior beliefs.
Namely, we hypothesise that (i) the most appropriate citation constituent is the the linear preferential attachment with relevance decay, \texttt{PA-RD}, (ii) the additive coupling between the constituents is more appropriate than the multiplicative one.
We will be able to judge whether there is strong evidence against these hypotheses based on the results of model selection.
For instance, if the more complex non-linear preferential attachment in the citation constituent is more prevalent, \texttt{PA-NL-RD} will be selected as the best model (as we will see, this is the case for some networks).
Or, we would find very large values of the characteristic relevance decay time \(\tau\) (of the order of, or larger than, the whole growth time), if the simplest preferential attachment without relevance decay would be a better choice for the citation constituent in a coupled model.

We proceed as follows: For each network, we estimate the parameters based on  5000 publications,  sampled uniformly at random,  which are added to the network during its growth, as explained in \cref{sec:computation}.
The exception is \gls{jrmp}, which has in total only 3006 publications, so we considered all of them for model fitting.
The number of citation edges \({|E^{pp}|}_S\) corresponding to the sample publications and used in likelihood calculation of \cref{eq:mle:ll:sample} varies between 4318 for \gls{jrmp} and 60131 for \gls{jhep}.

In \cref{tab:growth:mle-summary}, we present the selected models that have a relative likelihood  \(w_m \geq 0.01\) for each of the networks.
The table shows the estimates of the free parameters of the corresponding model, its relative likelihood \(w_m\) within the pool of the selected candidate models, and the log-likelihood of the model per considered edge.
We chose the latter instead of the total log-likelihood or the AIC score, because these strongly depend on the number of edges which differs across the studied networks. Even though we perform the likelihood estimation based on a fixed sample size of 5000 publications, the density of the networks differs.
Note that for \gls{jrmp}, three different models are listed because the model selection based on relative likelihood was inconclusive. However, in all of these three models the social constituent accounting for authors' number of publications is present.
The inconclusive results for \gls{jrmp} are likely due to the smaller size of \gls{jrmp} compared to the other networks.
For all other journals the data provided enough evidence to select one model over the other candidates.
\begin{table*}
	\caption{Parameter estimations for the selected growth models of nine physics journals. For the fixed parameters of the different models see \cref{tab:model_specification}. }\label{tab:growth:mle-summary}
	\centering\scriptsize
	\begin{tabularx}{\textwidth}{lf{3cm}cf{1cm}CCCCf{1cm}}
		\toprule
		\textbf{Journal} & \makecell{\textbf{Selected}\\ \textbf{Models}} & $\ln \mathcal{L} / {|E^{pp}|}_S$ & $w_m$ & $\alpha$ & $\beta$ & $\gamma$ & $\delta$ & $\tau$ \\\midrule
		\textbf{PR}                     &  PA-RD + NCOAUT        &  -4.00314 &  1.00   &  $0.90 (\pm0.020)$   &  ---                 &  ---                 &  $0.95 (\pm0.247)$ &  5185  \\\cmidrule(lr){1-9}
		\textbf{PRA}                    & PA-NL-RD              &  -4.13827 &  1.00   &  ---                 &  ---                 &  $1.14 (\pm0.224)$ &  ---                 &  8411  \\\cmidrule(lr){1-9}
		\textbf{PRC}                    & PA-NL-RD              &  -3.92745 &  1.00   &  ---                 &  ---                 &  $1.09 (\pm0.007)$ &  ---                 &  4860  \\\cmidrule(lr){1-9}
		\textbf{PRE}                    & PA-NL-RD              &  -4.14851 &  1.00   &  ---                 &  ---                 &  $1.21 (\pm0.013)$ &  ---                 &  9706  \\\cmidrule(lr){1-9}
		\multirow{3}{*}{\textbf{RMP}}   & PA-RD + NPUB          &  -2.84885 &  0.71   &  $0.85 (\pm0.124)$   &  ---                 &  ---                 &  $0.47 (\pm0.086)$ &   480  \\
		& PA-RD + MAXPUB        &  -2.84909 &  0.26   &  $0.84 (\pm0.016)$   &  ---                 &  ---                 &  $0.45 (\pm0.032)$ &   479  \\
		& PA-RD $\times$ NPUB   &  -2.84959 &  0.03   &  ---                 &  $0.19 (\pm0.021)$   &  ---                 &  $0.74 (\pm0.315)$ &   480  \\\cmidrule(lr){1-9}
		\textbf{JHEP}                   & PA-RD + NCOAUT        &  -3.60015 &  1.00   &  $0.85 (\pm0.007)$   &  ---                 &  ---                 &  $1.11 (\pm0.074)$ &  2828  \\\cmidrule(lr){1-9}
		\textbf{PR-HEP}                 & PA-RD + NPUB          &  -3.99631 &  1.00   &  $0.87 (\pm0.265)$   &  ---                 &  ---                 &  $0.65 (\pm1.403)$ &  7347  \\\cmidrule(lr){1-9}
		\textbf{Phys. Lett.}            & PA-RD + NPUB          &  -3.56096 &  1.00   &  $0.81 (\pm0.075)$   &  ---                 &  ---                 &  $0.40 (\pm0.186)$ &  2991  \\\cmidrule(lr){1-9}
		\textbf{Nuc. Phys.}             & PA-RD + NPUB          &  -3.61780 &  1.00   &  $0.85 (\pm0.129)$   &  ---                 &  ---                 &  $0.51 (\pm2.072)$ &  3278  \\
		\bottomrule
	\end{tabularx}
\end{table*}

\begin{figure*}
  \centering
  \includegraphics[width=.32\textwidth]{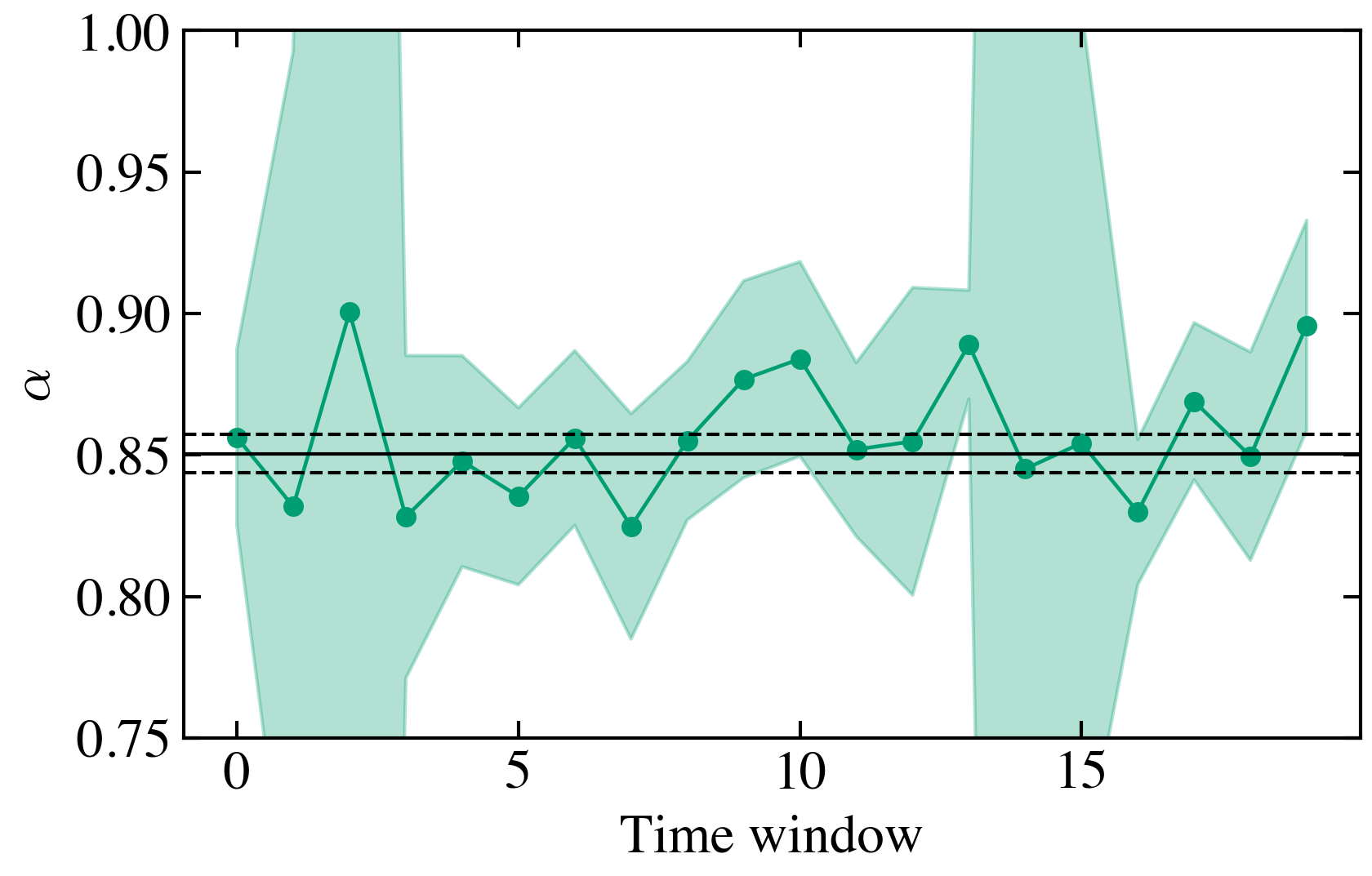}
  \hfill
  \includegraphics[width=.32\textwidth]{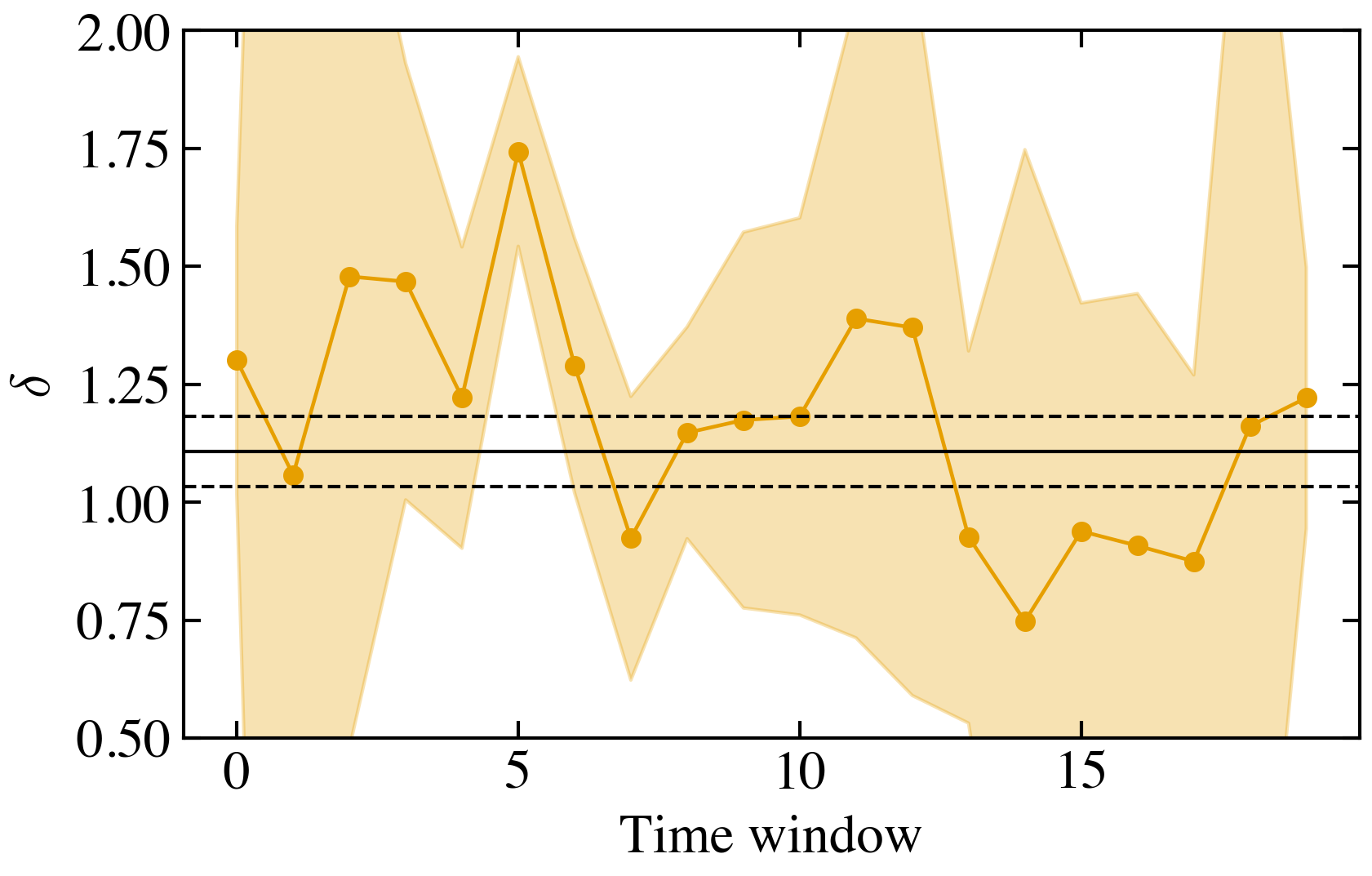}
	\hfill
  \includegraphics[width=.32\textwidth]{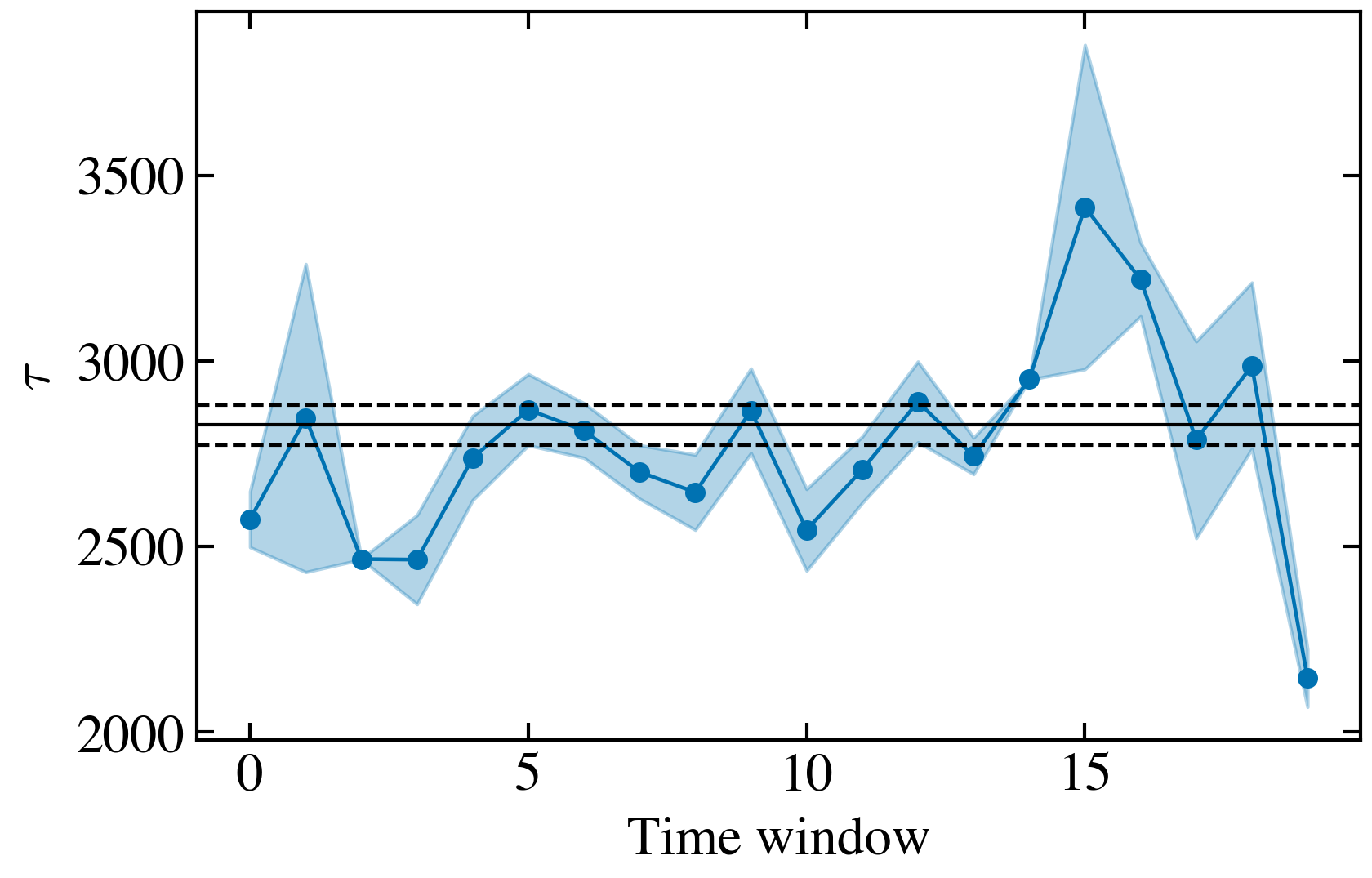}
  \caption{Model parameters of \texttt{PA-RD-NCOAUT} estimated for \gls{jhep} from  5000 sampled publications.
    The solid black lines correspond to the parameters and the region between the dashed lines  to their standard errors estimated on all 5000 sampled publications. The whole growth period ($x$-axis) was divided into 20 consecutive time windows, each comprising approximately 250 sampled publications.
The colored dots/lines indicate the parameters and the shaded area their standard errors within each time window. 
  }\label{fig:temp-trends}
\end{figure*}

Before discussing these results, we confirm that the model parameters are stable over the whole growth period of a network.
We illustrate this using the example of \gls{jhep}
Following the procedure described in \cref{sec:trends}, we first partition the whole growth period into twenty consecutive time windows and then estimate the parameters for the selected best model for this network, \texttt{PA-RD-NCOAUT}, based on each time window separately.
The results for the three parameters of the model are shown in \cref{fig:temp-trends}. 

One insight regards the relation between the estimates $\alpha^{T}$, $\delta^{T}$, $\tau^{T}$  based on the whole sample of 5000 publications (black lines) and the estimates  $\alpha^{R}$, $\delta^{R}$, $\tau^{R}$ based on the much smaller sub-samples for each time window. 
For the parameter \(\delta\), we see that the standard error for $\delta^{T}$ is within the standard errors for $\delta^{R}$ for all  but one estimates. 
Similarly, the standard error for $\alpha^{T}$ is within the standard errors for $\alpha^{R}$ for 18 out of
twenty estimates.
Hence, for these two parameters we can conclude that the MLE leads to a feasible outcome  and that the parameters are stable over time.

For the third parameter, \(\tau \), we also do not see any temporal trend in the estimates.
However, the standard errors for $\tau^{R}$ are unexpectedly small and their magnitude becomes  comparable to $\tau^{T}$.
This may be explained by the fact that $\tau$ encodes the characteristic \emph{time} of the relevance decay and therefore cannot be accurately estimated when publications are added within a relatively narrow time window.
We, hence, do not report the error estimates of  \(\tau \) in our results presented in \cref{tab:growth:mle-summary}.

The finding that, for the growth of the citation network, the parameter estimates are stable over time is not trivial because
our analysis of \gls{jhep} spans 20 years.
Between 1997 and 2017 the scientific landscape has changed considerably.
In particular, the pace of academic publishing has been increasing exponentially over the last decades~\cite{Lariviere2008}.
This accelerated growth in publications (and citations) does not affect our analysis because we have parametrised time in the network growth process by means of new publications.
When mapping the growth process to physical time, the earlier time windows in \cref{fig:temp-trends} span a much longer physical time interval compared to later time windows.
The fact that we achieve stable parameter estimates in our analysis supports the recent finding that attention decay in science is driven by the accumulation of newer publications rather than by the passing of time~\cite{Parolo2015}.

Going back to the results for the nine networks in \cref{tab:growth:mle-summary}, we find that model selection for most of the studied journals favors \emph{coupled models} of citation growth, i.e., models accounting for both a \emph{citation} and a \emph{social} constituent.
The citation constituent was best described by \texttt{PA-RD}, whereas for the social constituent different  candidates were chosen (mostly \texttt{NPUB}). 

The three exceptions are the journals \gls{jpra}, \gls{jprc} and \gls{jpre} published by the American Physical Society.
For these, model selection favors the non-linear preferential attachment, \texttt{PA-NL-RD}.
According to our research design presented in Table~\ref{tab:model_specification}, we excluded tests of \texttt{PA-NL-RD} together with social constituents.
Thus, we cannot argue about how \texttt{PA-NL-RD} would fare in coupled models.
But we already see that for \gls{jpra}, for which \texttt{PA-NL-RD} was selected, the exponent \(\gamma = 1.14 \pm 0.224\) is not significantly different from one, i.e., from the linear preferential attachment, \texttt{PA-RD}. 
Hence, we can suspect that in this case a coupled model does not give an advantage in the model selection, otherwise it had been selected. 

As a second insight, we note that model selection, for \emph{coupled models}, always favored an \emph{additive} coupling between citation and social constituents.
The weight of the social constituent measured by \((1 - \alpha)\), ranges between 0.07 and 0.19 across networks.

Only for \gls{jrmp} the selection of additive models is not strongly conclusive.
It lead to the largest relative likelihood $w_{m}= 0.71$ for the additive coupling, \texttt{PA-RD+NPUB}, but also
to a small, yet considerable, \(w_m = 0.03\) for its multiplicative counterpart, \texttt{PA-RDxNPUB}.

As a third insight, the social constituent \texttt{NPUB} based on the whole team of authors is always selected over the variant \texttt{MAXPUB} with only the most prominent author.
The exception is, again, \gls{jrmp}, where the model \texttt{PA-RD-MAXPUB} accounting for the highest number of publications among authors is selected with relative likelihood \(w_m = 0.26\), along with its counterpart that accounts for the number of publications of the whole team \texttt{PA-RD-NPUB} with \(w_m = 0.71\).

To summarise our findings, in most of the networks the \emph{coupled growth} model that includes a \emph{social constituent} has a higher likelihood to explain the observed citation dynamics than the simple growth model that only considers the previous history of the citation network.

\section{Conclusions}
\label{sec:growth:conclusion}

In this article, we have provided three contributions:
(i) a framework to model the \emph{growth} of networks consisting of \emph{different coupled layers}, 
(ii) a method for statistical parameter estimation and \emph{model selection} applicable to \emph{growing multi-layer networks}, and
(iii) a large-scale case study of the citation dynamics in nine different physics journals which demonstrates the applicability of our approach.

To model the growth of multi-layer networks is challenging because it combines two different dynamics, \emph{within} and \emph{between} network layers. 
The coupling between the layers usually does not allow us to separate the time scales for these dynamic processes.
Our modular approach provides a convenient way to cope with this in a stochastic manner.
Importantly, it allows to encode different \emph{hypotheses} about these dynamics, i.e. to generate different dynamic models, which can then be tested \emph{statistically} against empirical data.

This methodology requires two steps: (i) for each model, the parameters that match the data best have to be obtained using a Maximum Likelihood Estimation (MLE), and (ii) models with different complexity (i.e. number of free parameters) have to be compared in their ability to describe the data.
These two steps bear, again, a number of challenges that we address in this paper.
First, the MLE has to be applied to a growing network with coupled dynamic processes.
Here we followed a microscopic approach (see also \cite{Leskovec2008, Medo2014}), i.e. we focus on the temporal sequence of edges being added to the network.
Second, we need to ensure the temporal stability of the model, i.e.  the involved parameters should not change drastically over time. 
And third, the MLE needs to be computationally efficient, not only for computing parameters, but also for computing their standard errors.
This requires a scalable procedure that also works for large multi-layer networks. 
We solved this problem by means of sampling of growth events in the network.
Eventually, we need to provide an efficient procedure for \emph{model selection}, i.e. for comparing models of different complexity.
Here we have chosen relative likelihoods as the most insightful measure (see also \cite{Medo2014}).

In addition to the methodological contributions, our paper also provides interesting insights into scientometrics.
Specifically, we addressed the question to what extent social constituents, such as the number of \emph{previous} co-authors or publications,  impact the citation of publications.
This question was recast in the coupled growth of a  two-layer network, where one layer captures the publications and the second one the authors.
To construct such dynamic networks, we used data from nine physics journals, where each journal was captured in one two-layer network.
We proposed eleven models for the growth of the citation networks, which combine different forms for a citation and a social constituent.
These constituents represent different hypotheses about citation growth, mostly from the family of  preferential attachment models, and about social influences.

Our first insight regards the important role of the social constituent. 
We found that in the majority of the studied networks, the data is best described by \emph{coupled} models, i.e. models that  incorporate the coupling with the social constituent.
Among these, the additive coupling is selected in most of the cases, with the mixture weight of the social component between 7\% and 19\%.
Second, regarding the type of social influence, we learned that the total number of previous co-authors explains the citation data best, in most cases.  
This means, there is no strong statistical evidence for Merton's conjecture \cite{Merton1968} that the scientific community tends to cite a publication merely because of the most prominent author with the largest number of previous co-authors.
Third, it was interesting to note that not for all of the nine journals the citation dynamics was best described by the \emph{same} model.
There was clear evidence that preferential attachment with attention decay is the most promising candidate to explain the observed citations, which was already discussed in the literature \cite{Parolo2015}.
But three journals were better described by non-linear preferential attachment models, and there was also a variation in the form of the social influences.

Using the modeling framework provided in this paper, 
our analysis can be extended in different ways.
While our focus in this paper was on the growth of citations, 
the next step could incorporate a model component for authorship formation into the analysis.
Such a comprehensive model will facilitate further our understanding about the simultaneous co-evolution of citations and authorship relations.
It may shed more light on the feedback mechanisms between network layers and lead to insights about successful career paths of authors.

\appendix 

\section{MLE and model selection}\label{sec:app}

\begin{figure*}
  \centering
  \includegraphics[width=.32\textwidth]{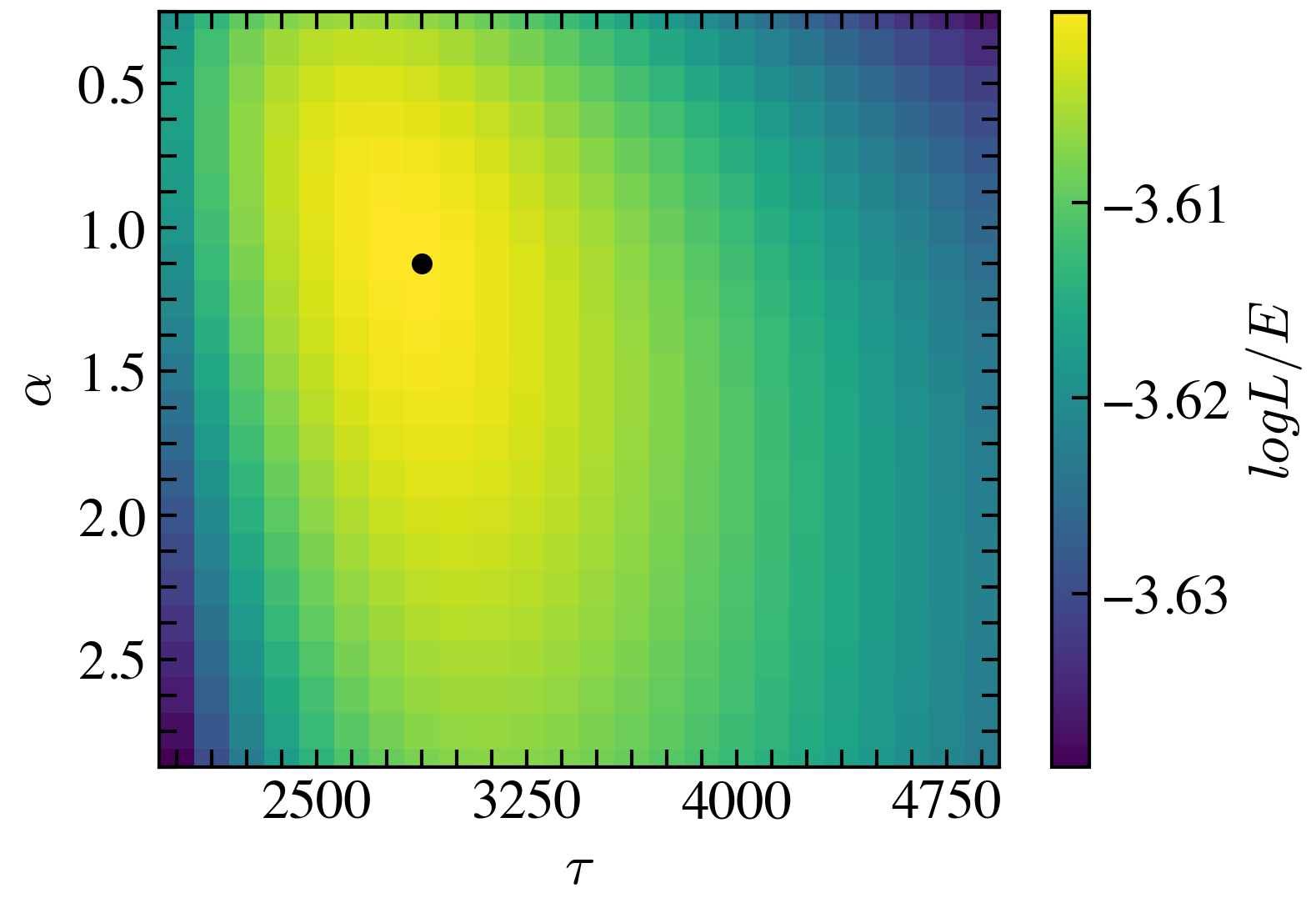}
  \hfill
  \includegraphics[width=.32\textwidth]{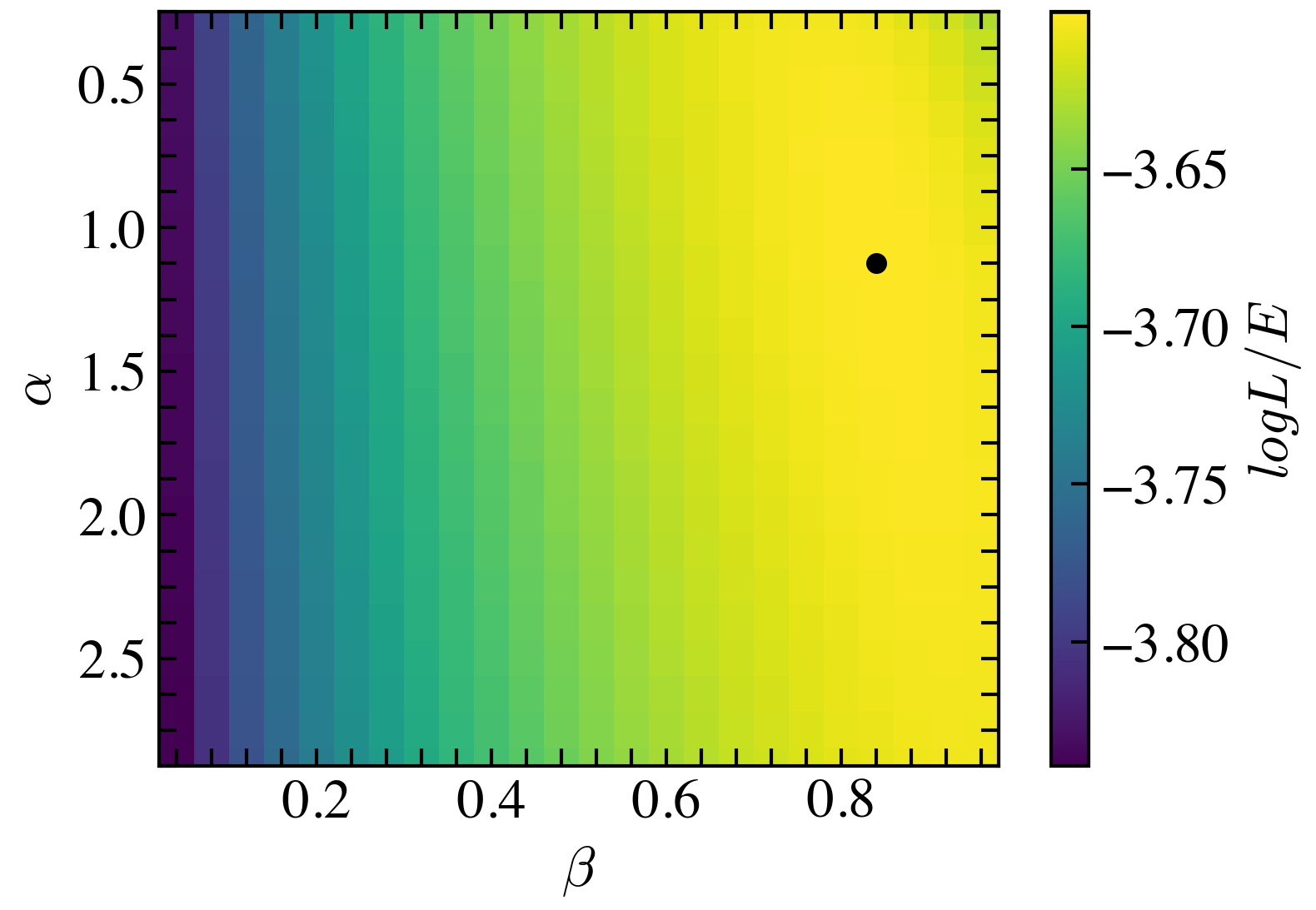}
  \includegraphics[width=.32\textwidth]{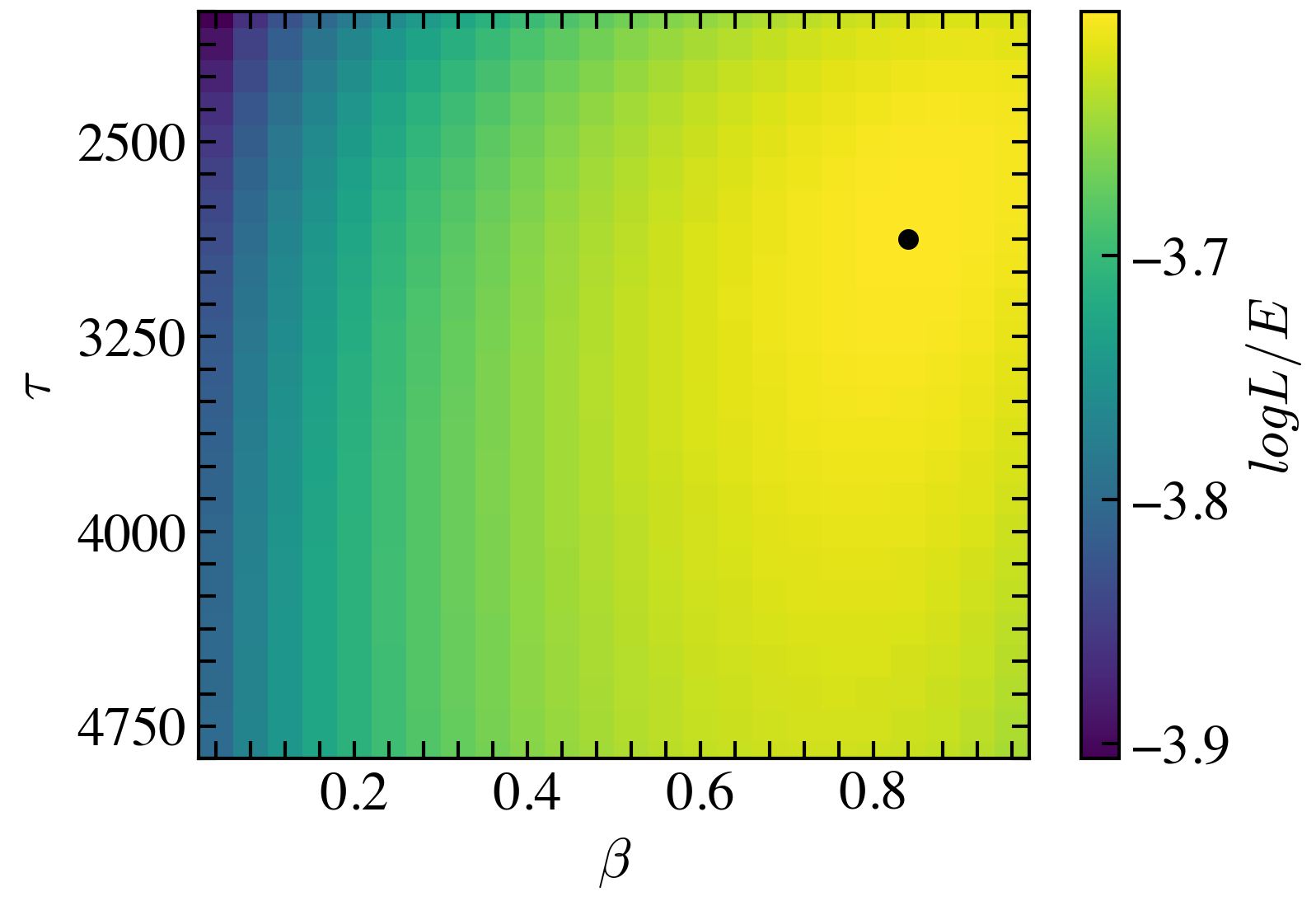}
	\caption{The log-likelihood function per edge for the model \texttt{PA-RD-NCOAUT} fitted to the network of \gls{jhep}. Each section plane of the two model parameters shown corresponds to the maximum likelihood value of the third parameter. The dot indicates the location of the maximum likelihood. It is the same in all three plots.}\label{fig:growth:grid}
\end{figure*}

In general, we cannot analytically calculate the log-likelihood function defined in \cref{eq:mle:ll,eq:mle:ll:sample}. 
The reason is that, e.g., even for the simplest linear preferential attachment, \cref{eq:PA}, the normalisation factors differ between summands \(\pi_{ij}^{\mu\nu}[G(t);\vecb{\theta}]\) for different \(\mu\nu\) and \(t\) in \cref{eq:mle:ll,eq:mle:ll:sample}.
Hence, we have to resort to numerical computation.
However, we can expect that the network growth models studied in this article and in the literature are described by a convex likelihood function with a unique maximum, as discussed in~\cite{Medo2014}.
We visually confirm this in \cref{fig:growth:grid} for one of our models that implements additive coupling between the citation and social constituents \texttt{PA-RD-NCOAUT}.

In \cref{sec:valid-synth-data}, we applied our method for the statistical evaluation of network growth models to synthetic networks with known growth mechanisms. 
This way we could confirm that the method correctly recovers the parameters of the true generating model.
In \cref{tab:growth:fits-sinthetic}, we present the results from estimating multiple growth models to such a network.
We see that the model selection based on relative likelihoods defined in \cref{eq:mle:rel-lik} correctly identifies the true model among multiple candidates.
\begin{table*}
	\caption{Model selection for a synthetic network defined in \cref{sec:valid-synth-data}.}\label{tab:growth:fits-sinthetic}
    \begin{tabularx}{\textwidth}{lCCCCCCC}
\toprule
     \textbf{Model} & \(\ln \mathcal{L} / {|E^{pp}|}_S\) & \(w_m\) & \(\alpha\) & \(\delta\) & \(\gamma\)  & \(\tau \) \\
\midrule
              PA & -2.74568 &    0.00 &         ---       & 2.86 \(\pm\) 0.170 &        ---          &  ---  \\
\textbf{PA-NAUT} & -2.74001 & \textbf{1.00} & 0.72 \(\pm\) 0.492 & 1.35  \(\pm\) 0.385 &   ---        &  ---  \\
           PA-RD & -2.74631 &    0.00 &         ---       & 2.33 \(\pm\) 0.123 &  ---  &  3760 \(\pm\) 11 \\
        PA-NL-RD & -2.75834 &    0.00 &         ---       &       ---            &   0.96  \(\pm\) 0.015  &  no convergence  \\
      PA-RD-NAUT & -2.75159 &    0.00 & 0.75 \(\pm\) 0.080  & 0.50 \(\pm\) 0.160 &  ---  &  no convergence  \\
   \bottomrule
    \end{tabularx}
\end{table*}

\section{Modelling details for one empirical network}\label{sec:jhep}

In \cref{tab:growth:mle-summary} we presented the results of model selection for nine empirical networks, for each considering a set of eleven candidate models.
Here, in \cref{tab:growth:fits-1213103}, we present the detailed outcome of fitting these eleven models to one of the networks, namely \gls{jhep}.
We report the resulting log-likelihood of the model per edge, i.e., \({\ln \mathcal{L}}/{{|E^{pc}|}_s}\), where \({|E^{pc}|}_s = 60131\) is the number of edges in the considered sample.
This means that each of the publications sampled for model evaluation cites on average  \({|E^{pc}|}_s/5000 \approx 12\) other publications \emph{within} the journal.
Recall that even though we use a relatively small sample of publications (about a third in case of \gls{jhep}, see \cref{tab:dm:sum:ih}), \emph{all} their citations to any publication in the whole network are considered.
\begin{table*}
  \caption{Fitting growth models to the network of \gls{jhep} based on a sample of 5000 publications that create \({|E^{pc}|}_s = 60131\) citations.}\label{tab:growth:fits-1213103}\scriptsize
    \begin{tabularx}{\linewidth}{lCCCCCCCC}
      \toprule
      \textbf{Model} & \(\ln \mathcal{L}/{|E^{pp}|}_S\) & \(AIC_m\) & \(w_m\) & \(\alpha\) & \(\beta\) & \(\gamma\) & \(\delta\) & \(\tau\)  \\
      \midrule
      UNIF             & -3.86236 &   464495 &  0.00  &  ---  &  ---  &  ---  &         ---        &  ---   \\
      PA               & -3.75232 &   451264 &  0.00  &  ---  &  ---  &  ---  & \(8.47 \pm 0.188\) &  ---   \\
      PA-RD            & -3.60539 &   433595 &  0.00  &  ---  &  ---  &  ---  & \(2.08 \pm 0.057\) &  2735  \\
      PA-NL-RD         & -3.61012 &   434164 &  0.00  &  ---  &  ---  & \(0.93 \pm 0.005\) &  ---   &  2644  \\
      \midrule
      PA-RD-NAUT       & -3.60267 &   433270 &  0.00 & \(0.82 \pm 1.000\) &  ---  &  ---  & \(0.77 \pm 1.000\)  &  2749  \\
      \textbf{PA-RD-NCOAUT} & -3.60015 &   432968 &  \textbf{1.00} & \(0.85 \pm 0.007\)  &  ---  &  ---   & \(1.11 \pm 0.074\)  &  2828  \\
      PA-RD-MAXCOAUT   & -3.60099 &   433068 &  0.00 & \(0.86 \pm 0.299\) &  ---  &  ---  & \(1.17 \pm 2.546\)   &  2786  \\
      PA-RD-NPUB       & -3.60229 &   433224 &  0.00 & \(0.90 \pm 0.012\) &  ---  &  ---  & \(1.42 \pm 0.437\)   &  2764  \\
      PA-RD-MAXPUB     & -3.60313 &   433325 &  0.00 & \(0.92 \pm 0.611\) &  ---  &  ---  & \(1.57 \pm 1.584\)   &  2779  \\
      \midrule
      PA-RDxNCOAUT     & -3.60167 &   433150 &  0.00 &  ---  & \(0.14 \pm 0.010\) &  ---  & \(2.38 \pm 0.471\)  &  2794  \\
      PA-RDxNPUB       & -3.60313 &   433326 &  0.00 &  ---  & \(0.10 \pm 0.006\) &  ---  & \(2.27 \pm 0.051\)  &  2740  \\
      \bottomrule
    \end{tabularx}
\end{table*}

First, we look at models of citation growth that implement only the citation constituent, i.e., no influences of social constituent capturing the author layer.
By comparing the log-likelihoods and the AICs, we find that the two models \emph{with} relevance decay, \texttt{PA-RD} and \texttt{PA-NL-RD}, considerably outperform the simple preferential attachment model, \texttt{PA}, which favors older nodes and, hence, does not reflect their decaying relevance. 

The two models \texttt{PA-RD} and \texttt{PA-NL-RD} estimate a characteristic relevance decay time \(\tau \approx 2700\).
Recall that we measure time in terms of newly added publications, so the value of \(\tau \) gives the number of publications that need to be added to the network, to reduce the relevance of a given publication by a factor \(e \approx 2.72\).
From the two models with relevance decay, \texttt{PA-RD} describes the data better.

The non-linear preferential attachment model, \texttt{PA-NL-RD}, indicates that  \(\gamma = 0.93 \pm 0.005 < 1\).
That means, the citation history matters less than expected from a simple preferential attachment model.
In \texttt{PA}, older publications have had more time to accumulate citations, hence tend to have higher in-degree only due to their higher age.
To compensate for that, we find from our parameter estimation that for \texttt{PA} the constant \(\delta=8.47 \pm 0.188\) has a very high value.
Given that $\delta$ increases the chance of being cited in particular for new publications, Eq.~(\ref{eq:2}), its high value also mitigates the relevance of older publications.

Next, we see that all models \emph{with} a social constituent perform better than models without it, reflected by the smaller values of both the log-likelihood and the AIC score. 
The model that best describes the data, with a relative likelihood \(w_m = 1.00\), is \texttt{PA-RD-NCOAUT}, which accounts for the total number of previous co-authors of the authors of the cited publication. 
The estimate of the mixture parameter \(\alpha = 0.85 \pm 0.007\) between the citation and the social constituents means that approximately 85\% of the received citations are driven by previously existing citations and  15\% are driven by properties of the authors, namely the counts of their of previous co-authors.

\section{Data}\label{sec:appendix_data}
\glsresetall

In the following, we describe the empirical data used in our analysis.
We used data from two sources, \gls{aps}~\cite{Note1} and INSPIRE~\cite{Note2}. We imported the data entries relevant for our study into a \texttt{Neo4j} graph database.
Storing the data as a graph with attributes attached to the nodes and edges allowed us to perform the needed constrained graph traversals very efficiently.
For  instance, for the model \texttt{PA-RD-NCOAUT} this data storage format allows to easily count paths of up to length three that traverse a certain sequence of node types (\texttt{publication} $\to$ \texttt{author} $\to$ \texttt{publication} $\to$ \texttt{author}) satisfying given constraints: the publications on the path must be in a given journal and the time stamps of subsequent publications on the paths must have earlier timestamps.

\paragraph*{\bf APS journals:}
The American Physical Society has been publishing physics journals since 1893.
From its foundation up to 1970, the \emph{Physical Review} journal published articles on all fields of physics.
The format of short letters was introduced in 1958 together with the journal \emph{Physical Review Letters} aiming to publish notable findings from all areas of physics in a rapid fashion.
In 1970 \emph{Physical Review} was split into four journals according to fields in physics.
Later splits and additions of new journals led to the existence of twelve \gls{aps} journals as of 2018.
From the set of historical and currently active journals, we studied the citation and authorship networks of  the following five journals: \gls{jpr}, \gls{jpra}, \gls{jprc}, \gls{jpre}, \gls{jrmp}.
This set covers a wide range of the physics sub-fields and types of publications. \gls{jrmp} is special in this selection as it focuses on review publications that consolidate the current state of a research topic instead of contributing original research outcomes.

\Gls{aps} freely provides the data on all their publications  \emph{``for use in research about networks and the social aspects of science''}.
For each publication, this data includes the title, the names and affiliations of the authors, the relevant time stamps (such as date received, date published), the journal in which it is published, and the list of other APS publications that it cites.
The major problem of the raw data for our analysis was that the authors are not uniquely identified.
I.e.,  the same name may describe multiple authors, or the same author may be represented in the data set by different name spelling, e.g., with the first name initial provided with one publication and the fully spelled first name with another publication.
To overcome this problem, we augmented the raw dataset with data on author disambiguation obtained using machine learning techniques~\cite{Sinatra2016}.
\Cref{tab:dm:sum:aps} summarises the networks of the five selected journals by the numbers of publications \(|V^{p}|\), authors \(|V^{a}|\), citations among publications \(|E^{pp}|\), and the authorship relations \(|E^{ap}|\).
	\begin{table}
		\centering
		\caption{Network summary for five journals published by the \gls{aps}.}
			\begin{tabularx}{\linewidth}{cCCCC}
				\toprule
				Network & \(|V^{p}|\) & \(|V^{a}|\) & \(|E^{pp}|\) & \(|E^{ap}|\) \\
				\midrule
				\gls{jpr}   &  46728  &  24307  &  253312  & 87386 \\
				\gls{jpra}  &  69147  &  41428  &  416639  & 144806 \\
				\gls{jprc}  &  36039  &  22672  &  253948  & 108844 \\
				\gls{jpre}  &  49118  &  36382  &  182701  & 95796 \\
				\gls{jrmp}  &  3006  &  3788  &  5282  & 5044 \\
				\bottomrule
			\end{tabularx}\label{tab:dm:sum:aps}
	\end{table}

	\paragraph*{\bf INSPIRE:}
This data set contains only publications relevant in the field of High-Energy Physics, thus allowing us to study how citation behaviour differs between journals on a single topic.
The underlying INSPIRE online information system is developed by a collaboration of CERN, DESY, Fermilab, IHEP, and SLAC.\@¸
It consolidates information on High-Energy Physics publications, researchers, collaborations and research data and allows access through a web interface, API and bulk downloads.
The major advantage of this data set is the high quality of author disambiguation, which is facilitated by personalised features, such as author profiles and paper claiming.

\begin{figure}[htpb]
    \centering
    \includegraphics[width=.45\textwidth]{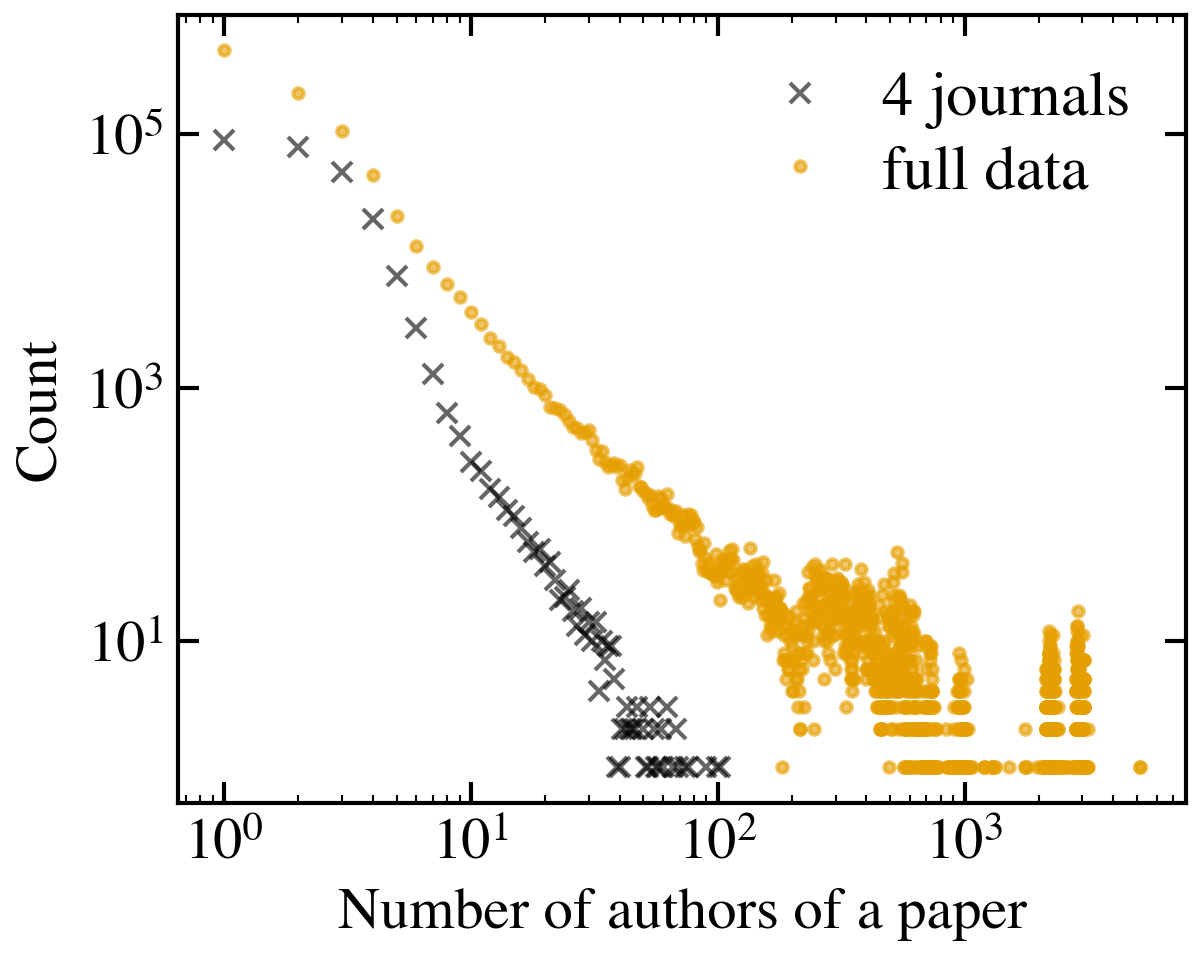}
    \caption{Distribution of the number of authors of a publication.  Yellow dots correspond to the full data set of INSPIRE, black `x' markers correspond to the sub-set from the three journals  \gls{jhep}, \gls{jpl}, \gls{jnp}, plus \gls{jpr_ins} used in our analysis.}\label{fig:dm:ih-numaut}
\end{figure}

High-Energy Physics differs from other fields in one crucial aspect: the significant number of publications with a huge number of authors.
As shown in \cref{fig:dm:ih-numaut}, the INSPIRE data set features 1278 publications with more than a thousand authors.
Based on the preceding data set SPIRES, it was pointed out~\cite{Newman2001} that the authors in High-Energy Physics have on average 173 co-authors, while in other fields the number ranges between 3.87 and 18.1.
Such publications with large numbers of authors relate mostly to collaborations, e.g. the \emph{ATLAS Collaboration}, on large-scale experiments, such as \emph{CERN-LHC-ATLAS}.

Publications with enormously long author lists are \emph{not} suited to study how the citation dynamics is influenced by social constituents, simply because of the dominating number of co-authors. 
Hence, we needed to focus a sub-set of the data to exclude such publications.
Specifically, from the INSPIRE data set we included publications identified as theoretical or general physics and excluded those identified as experimental,  based on corresponding tags present in the original raw data.
From the remaining sub-set of the INSPIRE data, we take the three journals that have the most, specifically more than 10,000, within-journal citations.
These are the journals \gls{jhep}, \gls{jpl}, \gls{jnp}, for which we constructed the three networks for our analysis. 
The distribution of the number of authors for these journals  is also shown in \cref{fig:dm:ih-numaut}.

\paragraph*{\bf Comparing APS and INSPIRE. }
The three networks obtained from INSPIRE are to be compared with a fourth network, \gls{jpr_ins}, which includes publications in all APS journals that were relevant enough in High-Energy Physics to be indexed in INSPIRE.
For comparison, we have applied the same selection criteria for  \gls{jpr_ins} as for the sub-set containing data from \gls{jhep}, \gls{jpl}, \gls{jnp}. 
Hence, studying the network from \gls{jpr_ins}, which also captures more than 10,000 citations, provides a different perspective on the citation dynamics in the APS journals.
\Cref{tab:dm:sum:ih} provides summary statistics for the four networks.
\begin{table}
    \centering
    \caption{Network summary for the four largest journals in the Inspire-HEP data set.}
    \begin{tabularx}{\linewidth}{lCCCC}
        \toprule
        Network & \(|V^{p}|\) & \(|V^{a}|\) & \(|E^{pp}|\) & \(|E^{ap}|\) \\
        \midrule
        \gls{jhep}  &  15739  &  7994  &  191990  & 39056 \\
        \gls{jpr_ins}  &  44829  &  33908  &  213625  & 115237 \\
        \gls{jpl}  &  22786  &  18078  &  56332  & 53089 \\
        \gls{jnp}  &  24014  &  18733  &  125252  & 60018 \\
        \bottomrule
    \end{tabularx}
    \label{tab:dm:sum:ih}
\end{table}

\bibliography{references}

\end{document}